# Chemistry-aware battery degradation prediction under simulated real-world cyclic protocols

**Authors:** Yuqi Li[1,2,4#], Han Zhang[3#], Xiaofan Gui[4#], Zhao Chen[1,2#], Yu Li[1,2#], Xiwen Chi[5], Quan Zhou[1], Shun Zheng[4*], Ziheng Lu[4*], Wei Xu[3], Jiang Bian[4], Liquan Chen[1,2], Hong Li[1,2*]

**Affiliations:**

[1] Key Laboratory for Renewable Energy, Beijing Key Laboratory for New Energy Materials and Devices, Institute of Physics, Chinese Academy of Sciences, Beijing, China

[2] College of Materials Science and Optoelectronic Technology, University of Chinese Academy of Sciences, Beijing, China

[3] Institute for Interdisciplinary Information Sciences, Tsinghua University, Beijing, China

[4] Microsoft Research

[5] International Center of Future Science, College of Chemistry, Jilin University, Changchun, China

[#]These authors contributed equally.

*Corresponding author. Email: shun.zheng@microsoft.com; zihenglu@microsoft.com; hli@iphy.ac.cn

**Abstract:** Battery degradation is governed by complex and randomized cyclic conditions, yet existing modeling and prediction frameworks usually rely on rigid, unchanging protocols that fail to capture real-world dynamics. The stochastic electrical signals make such prediction extremely challenging, while, on the other hand, they provide abundant additional information, such as voltage fluctuations, which may probe the degradation mechanisms. Here, we present chemistry-aware battery degradation prediction under dynamic conditions with machine learning, which integrates hidden Markov processes for realistic power simulations, an automated batch-testing system that generates a large electrochemical dataset under randomized conditions, an interfacial chemistry database derived from high-throughput X-ray photoelectron spectroscopy for mechanistic probing, and a machine learning model for prediction. By automatically constructing a polynomial-scale feature space from irregular electrochemical curves, our model accurately predicts both battery life and critical knee points. This feature space also predicts the composition of the solid electrolyte interphase, revealing six distinct failure mechanisms— demonstrating a viable approach to use electrical signals to infer interfacial chemistry. This work establishes a scalable and adaptive framework for integrating chemical engineering and data science to advance noninvasive diagnostics and optimize processes for more durable and sustainable energy storage technologies.

## Main

Understanding battery degradation is fundamental to the advancement of battery technology and the development of scalable, sustainable energy solutions[1]. Central to this effort are electrochemical curves, generated from detailed time-series data of battery charging and discharging, which provide indispensable insights into how batteries degrade over time[2,3]. These curves enable battery manufacturers to devise strategies to improve performance, extend lifespans, and optimize processes for recycling and cascading use. Typically, such data are collected under uniform testing protocols with standardized charge rates and environmental conditions[4-6], supporting reliable comparisons across battery models while enhancing parallel testing efficiency for large-scale data analysis.

However, real-world application of batteries presents a starkly different scenario[7-9]. Unlike controlled environment of laboratory tests, batteries in practical use are subjected to unpredictable operating conditions. Such randomness stems from fluctuating discharge powers and varying ambient temperatures, among other factors. These unpredictable loading conditions result in significant differences in batteries' cyclic behaviors compared to constant cyclic protocols.[10] Accompanying these randomized conditions is increased difficulty in predicting battery life. This randomness results in electrochemical curves that are far more irregular than those obtained from standardized tests. Such irregularities pose a significant challenge to existing predictive models, which are primarily designed to interpret data from uniform conditions[11]. As a result, these models often struggle to accurately predict battery behavior under the diverse and unpredictable scenarios encountered in real-world use, potentially impacting the reliability and safety of battery-powered devices. Compounding this issue is the incapability of current models to further correlating the degradation behavior with fundamental physiochemical features inside a battery[12]. While existing studies often train models using real-world usage data, these datasets typically do not provide direct insights into battery chemistry[7,13], thereby overlooking the detailed degradation mechanisms. Randomized battery cycling, despite its complex nature, provides unique opportunity to non-invasively probe the internal physiochemical features in a battery. The random pulses and power shift at different state-of-charge give abundant additional information such as relaxation times about batteries' internal functioning status[14]. However, how to process and leverage such signals to make predictions remains unknown.

To address these challenges, we developed a framework that explicitly links the irregular electrical signals with degradation behaviors and mechanisms to provide actionable insights to engineer durable and efficient battery systems. Firstly, we developed stochastic discharge protocols under desired driving conditions using hidden Markov processes (in this case highway driving condition from driver's real patterns[15]). By sampling from this process, we generated unique discharge protocols that align with the power distribution observed under real-world conditions. We developed an asynchronized scheduling system for batch testing batteries, enabling independent monitoring and parallel data collection under various operating conditions. Compared to directly collecting real user data, this approach offers more flexible and diverse approximations of real-world workloads, making it particularly well-suited for simulating and scaling actual scenarios under low-cost conditions, such as in a laboratory setting. Additionally, we developed a machine learning framework based on a polynomial-scale feature space composed from flexible operations on the early electrical signals. This feature space serves as the input to the machine learning models, leading to accurate and interpretable predictions regarding battery lifespan, critical knee points, and degradation. In particular, we examined 56 batteries post-cycling and conducted XPS on the graphite anode surfaces to gather information on the Solid Electrolyte Interphase (SEI) at various etching depths. This led to the database on battery interfacial chemistry, encompassing information on 8 kinds of surface elements. The XPS signals from these batteries naturally cluster into multiple groups, collectively reflecting six distinct failure mechanisms caused by varying interfacial chemistries. Using the same feature space generated from early electrical curves, we achieved accurate identification of these SEI patterns, establishing a link between non-invasively obtained electrical signals and internal chemical processes that typically require invasive methods. Such ‘chemistry-aware’ models predict interfacial chemistry for battery failure mechanisms. This means that while the model is primarily built on electrical signals, it is designed to output chemical information, making it fundamentally linked to battery interphase properties as a degradation mechanism probe.

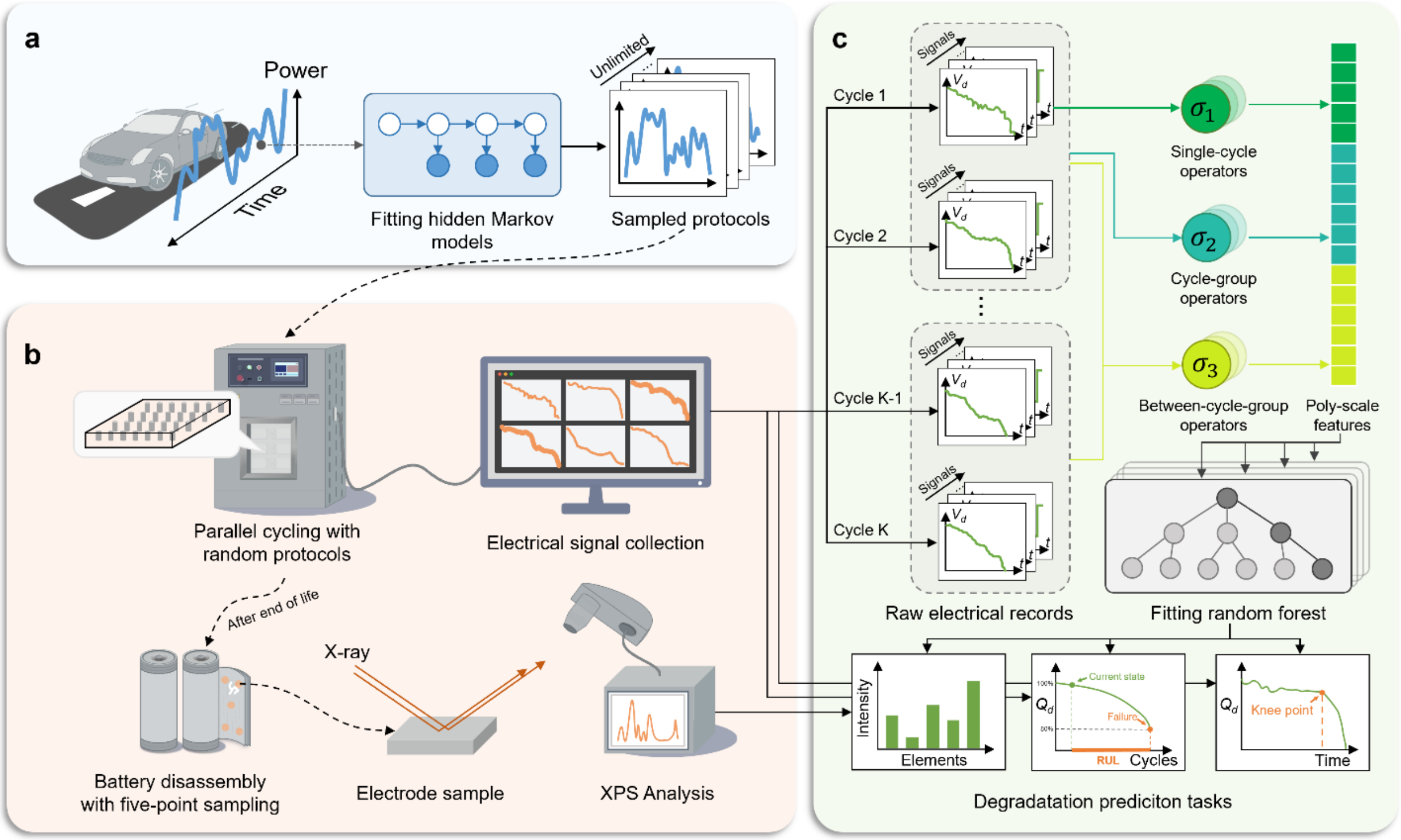

**Fig. 1: The overall workflow diagram of chemistry-aware battery degradation prediction under simulated real-world cyclic protocols.**

**a,** Protocol generation based on real-world power distributions, **b,** parallel data set collection including electrochemical curves and XPS signals (battery disassembly), **c,** polynomial-scale feature space construction and data-driven model training and inference.

## Simulating real-world battery cycling

In this study, we focus on electric vehicles (EVs) as a practical scenario for battery use[16]. To simulate real-world power variation in laboratory battery testers, we studied highway driving scenarios as defined by the U.S. Environmental Protection Agency's (EPA) US06 high-acceleration aggressive driving schedule[17]. Using vehicle dynamics models based on the Nissan Leaf[15], the EPA-defined speed curve was converted into a battery power curve (Supplementary Fig. 2). We train a hidden Markov model (HMM) to model this power distribution and predict future power needs. Using the learned HMM (Fig. 1a), we generated distinct discharge specifications based on real-world random power distribution curves. For each cycle, a unique workload was randomly selected, ensuring that every battery experienced a different configuration. In other words, the same battery will not always undergo the same cycling protocol; instead, it is subjected to continuously varying random cycles. We also examined the power distribution of these generated random discharge protocols and found them consistent with the observed power distribution of the US06 cycle (Fig. 2a).

To efficiently generate large-scale, randomized data following such real-world cycling protocols, we developed an automatic batch-testing system that integrates flexible cycling protocols with temperature control (Fig. 1b and Supplementary Figs. 3–5). This system uses a multi-tasking controller for parallel scheduling of battery testing channels, allowing for unmanned operation and rapid data production. The discharge protocols were designed to accommodate the complexity of real-world conditions while ensuring testing efficiency. Additionally, to further reduce testing time, we have implemented a two-stage constant power charging scheme—high power at low voltage followed by low power at high voltage—keeping charging times within an hour while addressing issues like Li plating[18]. Given the variable operating environments of EV batteries, we also included tests across seven different temperatures (from −10°C to 70°C) to capture the diverse conditions batteries might encounter.

In total, we collected charge-discharge cycles (136,737 in total) from 151 lithium-ion batteries with layered oxide cathode and graphite anode configurations (Fig. 3a). This database offers a more adaptable and varied representation of real-world battery usage compared to replaying the limited records from individual users[19,20]. A typical electric record collected under the simulated real-world cyclic protocol is shown in Fig. 2b. Compared to constant power discharge, the response voltage exhibits significant fluctuations due to the random input power protocol. The differences in cycling protocols lead to variations in battery lifespan distribution. Due to the high-speed highway conditions simulated, most batteries were in a rapid discharge state, resulting in lifespans or capacities generally shorter than the manufacturers' rated ones. As shown in Fig. 2d, simulated real-world cyclic protocols lead to a quicker and more unpredictable decrease in cycle life compared to constant power protocols, despite having the same average power (7 W).

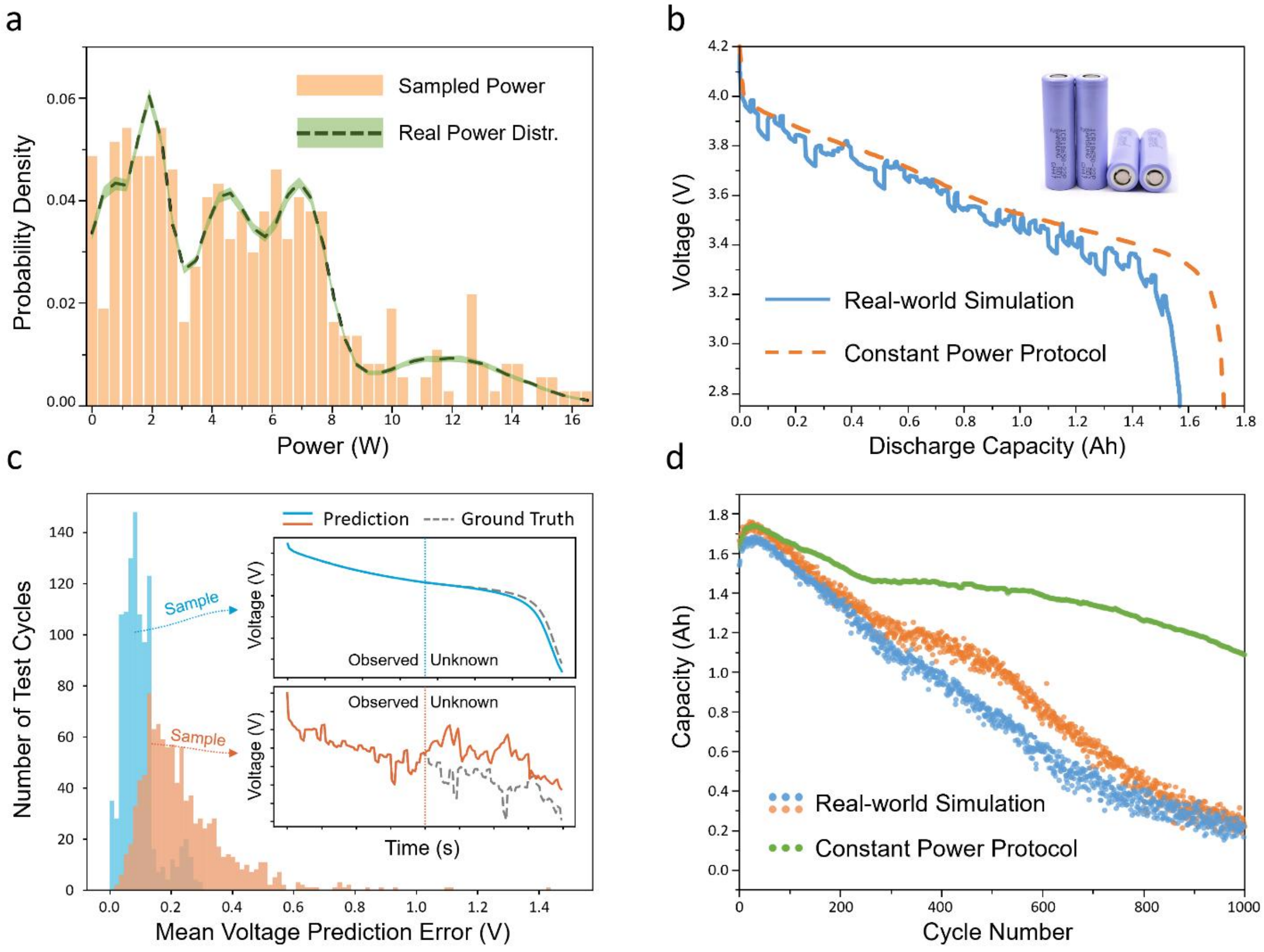

**Fig. 2. Simulating real-world cyclic protocols.**
**a**, Sample frequency of the simulated real-world cyclic protocols compared to the US06 protocol. **b,** Charge-discharge curves comparing the simulated real-world cyclic protocols and fixed cycling protocols. **c,** Voltage prediction error distribution of the constant current protocol and simulated real-world protocols. **d,** Comparison of battery degradation process under simulated real-world cyclic protocols (two kinds of varying power mode) and non-random conditions (constant power mode).

## Predicting battery lifespan and "knee point" under stochastic conditions

Predicting electrochemical behavior under the simulated realistic conditions is significantly more challenging than in the case of constant and regular discharge. First, the cyclic protocol is highly heterogeneous -- statistical analysis reveals that power transitions during single discharge cycles reach ~33.8 times in average, with frequent random power fluctuations further accentuating lifespan variability among batteries. However, most lifespan prediction studies to date have employed regular charge-discharge protocols, exemplified by the open-source database[4] containing 124 lithium iron phosphate batteries. Its single feature engineering approach based on changes in the discharge voltage curve ΔQ(V) is extensively used in both industry and academia for lifespan prediction in batteries. We observed a substantial performance degradation when adapting the classic ΔQ(V) feature method to random the current cycling protocols. As depicted in Fig. 3b and Supplementary Table 1, Mean Absolute Percentage Error (MAPE) is significantly high (75), indicating the inadequacy of the ΔQ(V) feature in our dataset. As an example, autoregressively predicting future voltage response is markedly more accurate under constant-current protocols, whereas prediction performance deteriorates substantially for irregular power inputs (Supplementary Fig. 6 and Fig. 2c).[21] Second, the dual influence of random operating conditions and temperature is more complex, leading to greater variability among batteries[22] (Supplementary Fig. 7). At lower temperatures (<10°C), batteries generally exhibit shorter lifespans due to impaired kinetic transmission and increased likelihood of lithium plating[23]. The optimal lifespan is attained at moderate temperatures (30-55°C), where elevated temperatures favorably enhance electrode reaction kinetics[24]. However, excessively high temperatures, such as 70°C, can destabilize the battery's thermodynamics[25]. Thus, the constructed random conditions, which include varying power levels and temperatures, make predicting aging states in batteries exceptionally challenging.

To enable effective degradation under randomized cyclic conditions, we took advantage of the abundant electric response signals originating from the abrupt power changes during cycling. Such power changes lead to discontinuity in voltage-time series and additional relaxations, making it unreliable to leverage human designed features as proposed in previous studies[4]. Therefore, we device a novel feature space construction based on statistical operations. By splitting the electrochemical curves and cycle into several groups and utilizing statistical operators to summarize the information (Fig. 1c), these features are iteratively refined by additional operators, creating a polynomial-scale feature space that is crucial for capturing complex patterns under random conditions. Detailedly, drawing from domain knowledge of lithium-ion batteries, we selected the capacity (Q), time (t), voltage (V), and current (I), energy (E), power (W) as seminal features, as these variables together imply the battery's electrochemical evolution. These variables lead to effective features like the well-known Q-V curves and less commonly explored ones like I-t curves (our realistic power mode results in corresponding current fluctuations). For each variable, we first split the cycles into several groups and divided the electric curves into multiple segments for each cycle. By splitting the cycles and the raw electric signals into segments, our method can capture local information which may get easily overlooked globally. On the one hand, as we split the signals into smaller pieces, in the limit the generated features will contain every detail in the raw data, yielding an extremely high dimensional feature space. On the other hand, if we avoid the split completely and consider exactly the $10^{th}$ and $100^{th}$ cycle voltage-capacity curves, our model degrades to the method proposed by Severson et al[3]. In practice, we adopt seven cycle groups and four signal segments to balance complexity and performance, as shown in Supplementary Fig. 8. After the split, we applied different extractors on these segments to extract statistical descriptors within each cycle, within each cycle group, or between different cycle groups. Note that these statistical descriptors are extracted without manual intervention and contain rich and fine-grained information of battery degradation under realistic workloads. We further enriched the features by iteratively applying statistical operators (including aggregation and scaling operators, detailed in Supplementary Table 2) on these extracted descriptors, leading to a polynomial-scale space up to 112,900 features including 56,450 discharging features and 56,450 charging features.

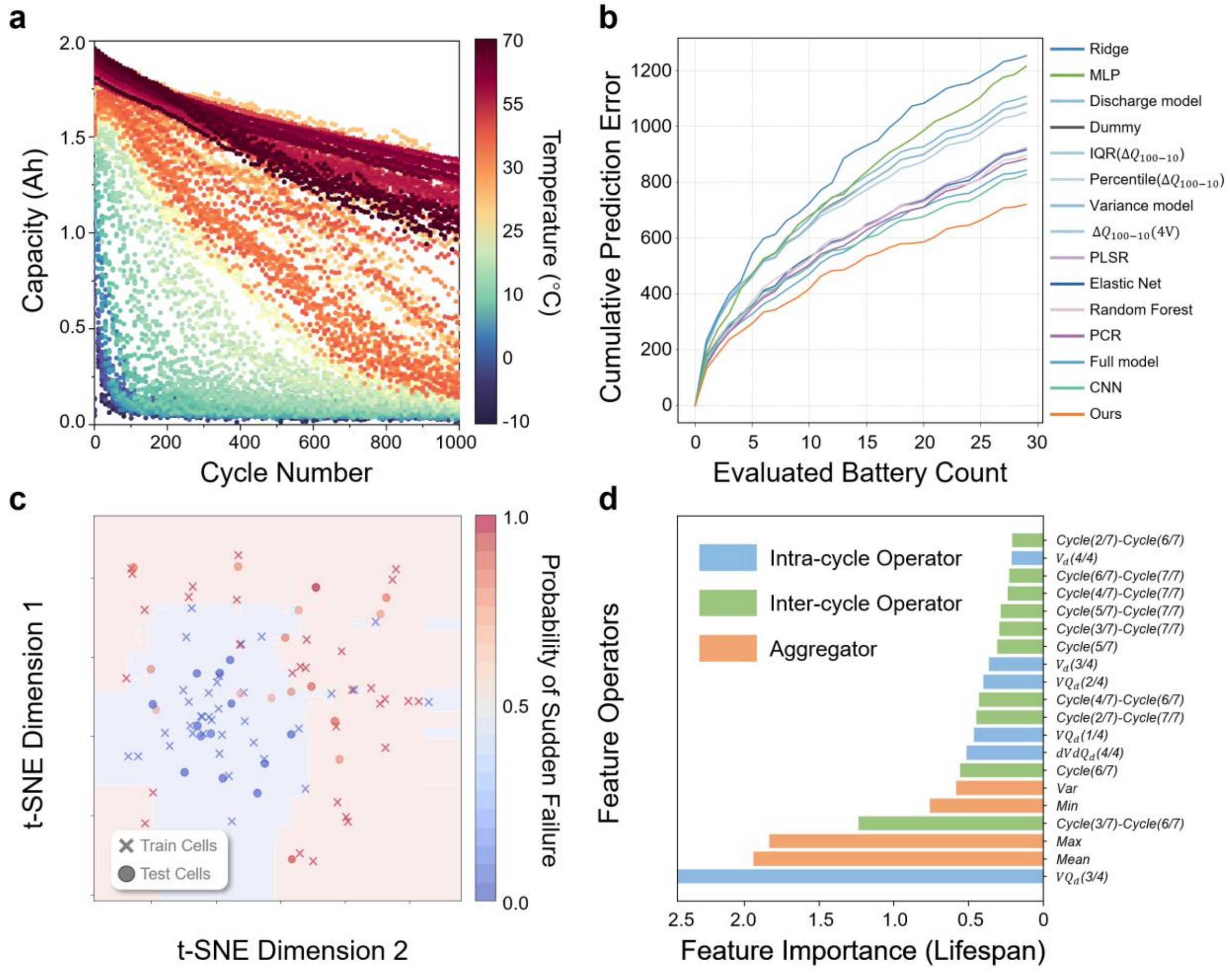

**Fig. 3. Prediction of Battery Remaining Life and "Knee Point" under stochastic cyclic protocols.**

**a**, Discharge capacity for the first 1,000 cycles of 18650-type cells collected by simulated real-world cyclic protocols. The related battery labels have been shown in Supplementary Fig. 17. **b**, Cumulated mean absolute prediction error averaged over 16 seeds, for various models as a function of the number of test data points. The sequence of adding test battery cells are identical among these methods. **c**, The t-SNE visualization of the model prediction and learned decision boundary forecasting rapid capacity drops. **d**, The most salient operators ranked by the importance of the associated features in prediction battery lifespan.

Based on the feature space, we employed a random forest regression model (Fig. 1c) to predict battery lifespan (80% of cell nominal capacity). A random forest consists of a bunch of decision trees that focus on a subset of features and samples. The outputs of these decision trees are aggregated through voting to yield a robust prediction. Such a forest model balances both predictive performance and interpretability in forecasting. In this research, we utilized data from the initial 50 cycles for prediction. We excluded 56 batteries with fewer than 50 cycles at 80% lifespan to ensure experimental rigor, retaining 95 suitable batteries for model training and evaluation. The model achieves superior prediction power on such randomized cyclic conditions, as shown in Supplementary Table 1. Notably, all the comparison methods relied on manually designed features, whereas our model demonstrated higher adaptability. In Fig. 3b, our model exhibits the best performance with a stable and minimal slope, indicating that the prediction error is consistent across all batteries without any extreme outliers. In contrast, the other methods[4,26] fall into three distinct tiers: models from Variance model to Convolutional Neural Network (CNN) show similar performance, followed by models from the Discharge model to Percentile ($\Delta Q_{100-10}$), with Ridge Regression and Multilayer Perceptron (MLP) performing the worst. A key advantage of the current Random Forest model is its inherent ability to assess feature importance, measured by the total information gain contributed by each feature across all decision trees. Fig. 3d and Supplementary Table 3 highlight the feature importance scores of the operators used in constructing our feature space. In the lifespan prediction task, the third segment of the voltage-capacity curve during discharge is identified as the most important feature. This is likely due to the significant capacity variation within the 3.5 V to 3.87 V range (Supplementary Fig. 7), which captures signals related to capacity decay and lifespan prediction[27]. This voltage range corresponds to the redox processes of key transition metals like nickel, essential for accurately predicting battery lifespan. Additionally, the selected cycle groups are not the earliest or latest in the cycle sequence. We hypothesize that this is because early electrochemical processes tend to be more stable with slower aging, while mid-cycle aging—such as SEI growth—leads to more noticeable performance changes[28].

We continued using the Random Forest model as a classifier to predict whether a battery will show a knee point during the cycling (Supplementary Fig. 9), i.e., abrupt change from slow to fast degradation. By employing the same feature space, we achieved an accurate prediction of the existence of knee point, demonstrating the generalizability of our feature space across different aging-related tasks (Supplementary Fig. 10 and Supplementary Table 4). This success can be attributed to the rich information encapsulated in our automatically generated feature space. To further illustrate the knowledge learned by the model from this feature space, we projected the high-dimensional battery features into a 2D space, as shown in Fig. 3c. Different markers are used to distinguish between training and test samples, with the model's decision boundary clearly displayed. It is evident that the separation learned by the model is highly robust, effectively distinguishing between batteries that will and will not show a knee point during degradation in both the training and test sets.

### Predicting Battery Interfacial Chemistry and Failure Mechanisms under simulated real-world cycling

While we have successfully utilized early electrical signals to accurately predict a battery's remaining life and knee point, a deeper understanding of the fundamental causes of battery degradation is necessary as such knowledge is key to enable targeted strategies to minimize degradation and enhance battery longevity. To enable this, we leverage X-ray and optical probes to gather abundant interfacial chemistry data by disseminating degraded batteries and classify the degradation mechanisms into six categories. We correlate such degradation mechanisms with non-invasive early battery cycling signals under randomized conditions by predicting the elemental distribution in the dataset using the machine learning model based on the feature space.

In this work, we majorly used X-ray Photoelectron Spectroscopy (XPS) to characterize the anode SEI (Fig. 1b) to gather interfacial chemistry data, as battery failures are usually more severe on the anode side and primarily occur at interfaces (Fig. 4a)[1]. This is mainly because SEI layer forms on the graphite anode surface during the initial charging to provide kinetic protection to the electrode[29]. During cycling, non-uniform ion flux can cause the SEI to rupture, leading to further electrolyte reactions and thickening. The uneven SEI can easily induce uneven surface electric fields, leading to lithium plating. A reasonable hypothesis is that by analyzing the nature of the SEI, we may be able to provide a quantitative failure analysis. The sample stage of the XPS equipment can accommodate about 20 samples of approximately 1 $cm^2$ each at a time, meeting the needs for high-throughput testing. Considering the inconsistency in different areas inside the battery, we adopted a five-point sampling method commonly used in biology. We collected electrode slices from five different positions (Fig. 1b) and included an etching step to capture deeper SEI information and clean the surface impurities. This resulted in the battery interfacial chemistry database using simulated real-world cyclic protocols, including 552 total XPS spectra (pre-etching and post-etching) from 56 batteries.

As shown in Supplementary Fig. 12, the SEI is primarily composed of non-metal elements such as carbon (C), oxygen (O), fluorine (F), and phosphorus (P). Further selecting different binding energy positions, we can also extract information on metal elements like lithium (Li), nickel (Ni), cobalt (Co), and manganese (Mn). It is evident that batteries with similar lifespans, but different SEI compositions likely have different failure mechanisms. Although XPS data can quantitatively depict interfacial chemistry, the numerous types of elements and complex varying proportions make it difficult to directly correlate the specific element content with battery aging mechanisms. Instead, we use cluster analysis to examine the interfacial chemical patterns based on XPS data obtained after etching. Interestingly, we obtained six distinct clustering patterns, which could cover almost all the XPS data (Supplementary Table 5). This means that the element proportions corresponding to the six cluster centers represent the primary types of interfacial chemistry post-failure. These six types of interfacial chemistry patterns are displayed along with the corresponding average operating temperatures and average lifespans of the batteries in Fig. 4b and Supplementary Fig. 13:

**Pattern 1** (denoted as LT-SL: low temperature and short cycling life), characterized by the highest lithium content in the SEI (~32.5%) and the shortest average lifespan (~75 cycles), combined with the lowest average temperature (~20°C) and fast-charging cycles, suggests a severe lithium plating mechanism. Additionally, this pattern has the highest transition metal ion crossover (Ni+Co+Mn ~3.1%), possibly indicating short-circuiting between plated lithium and the cathode. Scanning electron microscopy (SEM) observations confirmed severe lithium plating on the graphite surface for this pattern (Supplementary Fig. 14).

**Patterns 2** (denoted as MT-MLL: medium temperature and moderately long cycling life) and **Pattern 3** (denoted as MT-SL: medium temperature and short cycling life), with higher average temperatures (~28°C), show a reduced risk of lithium plating[30] (as indicated by the decreased Li signal), but they have significantly different average lifespans (~216 vs. ~166 cycles). This discrepancy is due to their different interfacial chemistries: MT-MLL's abnormally high O signal (~31.8%) likely originates from an SEI rich in lithium oxide, which can inhibit dendrite growth[31]. In contrast, MT-SL's high C signal (~30.6%) implies a higher proportion of porous organic components, making it more susceptible to lithium dendrite penetration[32].

**Pattern 4** (denoted as MT-ML: medium temperature and medium cycling life), with a similar average temperature (~30°C) as MT-MLL and MT-SL, shows an improved cycle life. Its interfacial chemistry includes higher P (~3.6%) and F (~17.5%) signals, indicating an SEI rich in LiF and $Li_3PO_4$, which provides lower electronic conductivity and higher ionic conductivity for a more stable SEI[33]. Despite this, Patterns 2-4 still show high lithium content, suggesting a significant possibility of lithium plating. SEM observations (Supplementary Fig. 15) confirmed lithium plating along with noticeable SEI thickening, indicating a mixed aging mechanism involving both lithium plating and SEI growth.

**Pattern 5** (denoted as HT-LL: high temperature and long cycling life) arises with further increased average cycling temperatures above 50°C, where observed a significant improvement in lifespan (>300 cycles).

**Pattern 6** (denoted as HT-LRL: high temperature and longer cycling life), compared to HT-LL, is characterized by higher P+F signals (~26.76% vs. ~19.29%) and lower C+O signals (45.52% vs. ~57.43%), features an SEI that is rich in inorganic components and poor in organic components. This type of SEI provides slower interfacial thickening, delaying the degradation of lifespan. Although this pattern has a higher concentration of transition metal ions on the surface (~2.25%), the dense SEI can prevent internal electrode crosstalk[34]. This confirms that at higher temperatures, the aging mechanism is more related to SEI growth rather to transition metal ion dissolution. SEM observations further confirmed that this pattern primarily involves SEI growth (Supplementary Fig. 14).

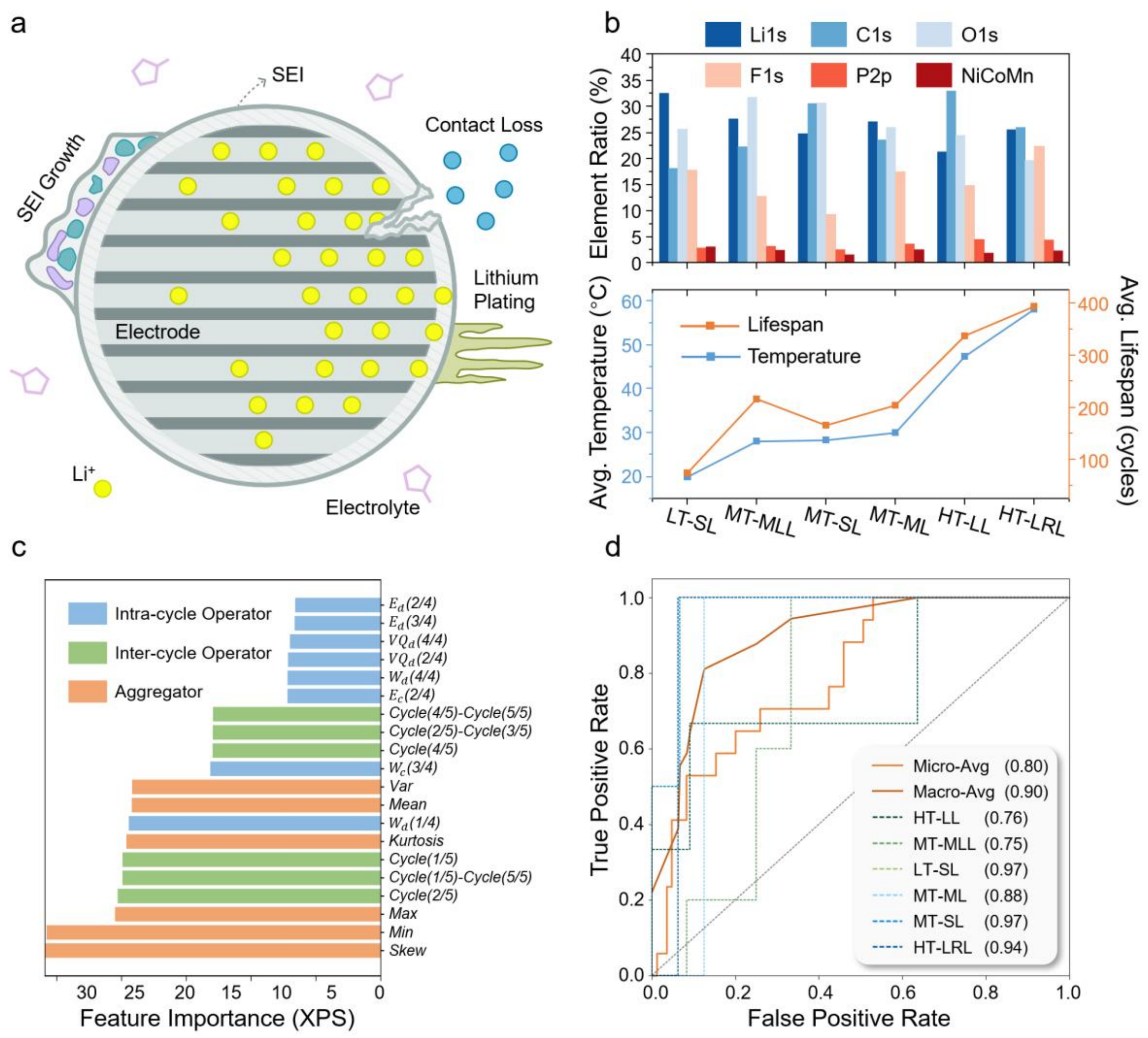

**Fig. 4. Prediction of Battery Interfacial Chemistry after Failure under simulated real-world cyclic protocols.**

**a**, Diagram of battery interface failure. **b**, Six types of XPS patterns identified by clustering the XPS element densities, along with detailed average elemental ratios, average operating temperature, and average remaining useful life of the batteries. **c**, The most salient operators ranked by the importance of the associated features in prediction battery lifespan. **d**, Receiver Operating Characteristic (ROC) curves for the interfacial chemistry pattern recognition. The average performance and per-pattern identification capability are provided, along with the corresponding area under the curve (AUC) scores in the legend.

Optical imaging (Supplementary Fig. 16) was further carried out to further confirm the above SEM images and XPS results. Therefore, our method indicated a potential pathway for quantifying failure mechanisms through high-throughput XPS analysis combined with battery disassembly. Specifically, beyond the traditional models of lithium plating and SEI growth (though closely related), our findings identify six refined patterns (with different proportions of elements in the XPS) corresponding to various interfacial chemistry aging mechanisms.

To correlate the failure mechanisms with the battery cyclic signals under simulated real-world conditions, we use the automatically generated polynomial-scale feature space as the input for the machine learning model. The 20 most salient features are shown in Fig. 4c and Supplementary Table 6. We observed that these features differ significantly from those associated with battery degradation trajectories. The earliest cycles were successfully identified, as the initial electrical signals often determine the composition of the SEI layer. While features during discharge may be important for predicting battery life and identifying knee points, features during charging could be more critical for predicting failure mechanisms, as the charging process often leads to the regeneration of the SEI layer or lithium plating. Crucially, the random forest employed for mechanism prediction (six XPS patterns) yielded small prediction errors, as depicted in Fig. 4d, demonstrating a robust capability in discriminating different aging scenarios as well as analyzing different elemental ratios. This implies that it is feasible to predict SEI composition directly from early electrical signals, enabling the inference of specific failure mechanisms without the need for intrusive battery disassembly processes. By establishing a strong correlation between electrical signals and interfacial chemical signals, current machine learning model could advances the prediction of both degradation trajectories and failure mechanisms.

## Conclusion

We developed a scalable, data-driven approach to model battery degradation under randomized discharge protocols, revealing distinct degradation patterns compared to regular protocols. By using hidden Markov processes to simulate real-life power workloads and an automated controller for batch operation, we collected life-cycle data from 151 batteries, each cycle employing a unique random discharge protocol. Utilizing this dataset, we divide and extract various types of electrical signals from early cycles and automatically arrange and compose them to obtain a polynomial-scale feature space. This enabled accurate predictions of battery lifespan and knee points of battery failure. Furthermore, after dissecting 56 cells after cycles, we analyzed graphite anode surfaces via XPS to gather SEI data at various etching depths, creating a battery interfacial chemistry database. Incorporating temperature, lifespan data, SEM, and optical analysis, we discovered that interfacial chemistry information can be automatically clustered into six patterns based on their correlations. These patterns represent six distinct aging mechanisms, which surpass the simplistic models of lithium plating and SEI growth. By leveraging the automatically generated feature space, we accurately linked electrical signals to interfacial chemical compositions, offering a robust framework for predicting failure mechanisms.

This work presents an integrated and highly automated pipeline for battery testing and analysis towards real-world applications, encompassing protocol generation, data collection, feature space construction, and machine learning-based aging prediction. While using dynamic driving conditions as a case study, the approach is broadly applicable across diverse scenarios beyond highways. The scalability of the system stems from its automation, which enables the simulation of real-world working conditions and the discovery of rich, previously untapped voltage-current responses under randomized protocols. These responses open new opportunities for feature learning, allowing accurate predictions of complex tasks, including even interfacial chemical information. By integrating chemical engineering principles like system modeling and process scalability with machine learning, this framework provides a non-destructive, adaptable method for diagnosing and predicting degradation. However, since our dataset is constrained to a single battery chemistry under controlled conditions, further studies incorporating a wider range of battery types and real-world environmental variations—such as temperature fluctuations, rest periods, and calendar aging—are needed to validate its broader applicability. Moreover, while the model successfully links early-cycle electrochemical signals to interfacial chemistry, future work could refine these predictions by integrating additional chemical characterization techniques, providing deeper insights into degradation mechanisms.

Furthermore, this study lays the foundation for standardized testing protocols under realistic conditions and offers insights into aging mechanisms to guide future battery optimization. Beyond batteries, this methodology—particularly the automatically constructed polynomial feature space—can be applied to other scientific domains characterized by complex, dynamic system behaviors under stochastic conditions, offering a blueprint for scalable diagnostics and optimization.

## References

1. Palacín, M.R. & de Guibert, A. Why do batteries fail? *Science* **351**, 1253292 (2016).
2. Attia, P.M., *et al.* Closed-loop optimization of fast-charging protocols for batteries with machine learning. *Nature* **578**, 397-402 (2020).
3. Ng, M.-F., Zhao, J., Yan, Q., Conduit, G.J. & Seh, Z.W. Predicting the state of charge and health of batteries using data-driven machine learning. *Nature Machine Intelligence* **2**, 161-170 (2020).
4. Severson, K.A., *et al.* Data-driven prediction of battery cycle life before capacity degradation. *Nature Energy* **4**, 383-391 (2019).
5. Roman, D., Saxena, S., Robu, V., Pecht, M. & Flynn, D. Machine learning pipeline for battery state-of-health estimation. *Nature Machine Intelligence* **3**, 447-456 (2021).
6. Zhang, H., *et al.* Battery lifetime prediction across diverse ageing conditions with inter-cell deep learning. *Nature Machine Intelligence* (2025).
7. Wang, Q., *et al.* Large-scale field data-based battery aging prediction driven by statistical features and machine learning. *Cell Reports Physical Science* **4**(2023).
8. Steininger, V., Rumpf, K., Hüsson, P., Li, W. & Sauer, D.U. Automated feature extraction to integrate field and laboratory data for aging diagnosis of automotive lithium-ion batteries. *Cell Reports Physical Science* **4**(2023).
9. Moy, K., Ganapathi, D., Geslin, A., Chueh, W. & Onori, S. Synthetic duty cycles from real-world autonomous electric vehicle driving. *Cell Reports Physical Science* **4**(2023).
10. Geslin, A., *et al.* Dynamic cycling enhances battery lifetime. *Nature Energy* (2024).
11. Li, Y., *et al.* Data-driven health estimation and lifetime prediction of lithium-ion batteries: A review. *Renewable and sustainable energy reviews* **113**, 109254 (2019).
12. Geslin, A., *et al.* Selecting the appropriate features in battery lifetime predictions. *Joule* **7**, 1956-1965 (2023).
13. Lyu, D., Zhang, B., Zio, E. & Xiang, J. Battery cumulative lifetime prognostics to bridge laboratory and real-life scenarios. *Cell Reports Physical Science* **5**(2024).
14. Ciucci, F. & Chen, C. Analysis of electrochemical impedance spectroscopy data using the distribution of relaxation times: A Bayesian and hierarchical Bayesian approach. *Electrochimica Acta* **167**, 439-454 (2015).
15. Yang, X.-G., Liu, T. & Wang, C.-Y. Thermally modulated lithium iron phosphate batteries for mass-market electric vehicles. *Nature Energy* **6**, 176-185 (2021).
16. Liu, W., Placke, T. & Chau, K.T. Overview of batteries and battery management for electric vehicles. *Energy Reports* **8**, 4058-4084 (2022).
17. Smith, K. & Wang, C.-Y. Power and thermal characterization of a lithium-ion battery pack for hybrid-electric vehicles. *Journal of power sources* **160**, 662-673 (2006).
18. Gao, T., *et al.* Interplay of lithium intercalation and plating on a single graphite particle. *Joule* **5**, 393-414 (2021).
19. Lombardo, T., *et al.* Artificial Intelligence Applied to Battery Research: Hype or Reality? *Chemical Reviews* **122**, 10899-10969 (2022).

20. Klass, V., Behm, M. & Lindbergh, G. A support vector machine-based state-of-health estimation method for lithium-ion batteries under electric vehicle operation. *Journal of Power Sources* **270**, 262-272 (2014).
21. He, W., Williard, N., Osterman, M. & Pecht, M. Prognostics of lithium-ion batteries based on Dempster–Shafer theory and the Bayesian Monte Carlo method. *Journal of Power Sources* **196**, 10314-10321 (2011).
22. Ma, S., *et al.* Temperature effect and thermal impact in lithium-ion batteries: A review. *Progress in Natural Science: Materials International* **28**, 653-666 (2018).
23. Zhang, N., *et al.* Critical review on low-temperature Li-ion/metal batteries. *Advanced Materials* **34**, 2107899 (2022).
24. Yang, X.-G., Zhang, G., Ge, S. & Wang, C.-Y. Fast charging of lithium-ion batteries at all temperatures. *Proceedings of the National Academy of Sciences* **115**, 7266-7271 (2018).
25. Li, Y., *et al.* Ultralow-concentration electrolyte for Na-ion batteries. *ACS Energy Letters* **5**, 1156-1158 (2020).
26. Attia, P.M., Severson, K.A. & Witmer, J.D. Statistical learning for accurate and interpretable battery lifetime prediction. *Journal of The Electrochemical Society* **168**, 090547 (2021).
27. Li, W., Song, B. & Manthiram, A. High-voltage positive electrode materials for lithium-ion batteries. *Chemical Society Reviews* **46**, 3006-3059 (2017).
28. Liu, T., *et al.* In situ quantification of interphasial chemistry in Li-ion battery. *Nature Nanotechnology* **14**, 50-56 (2019).
29. Li, Y., Lu, Y., Adelhelm, P., Titirici, M.-M. & Hu, Y.-S. Intercalation chemistry of graphite: alkali metal ions and beyond. *Chemical Society Reviews* **48**, 4655-4687 (2019).
30. Li, Y., *et al.* Origin of fast charging in hard carbon anodes. *Nature Energy* **9**, 134-142 (2024).
31. Kim, M.S., *et al.* Suspension electrolyte with modified Li+ solvation environment for lithium metal batteries. *Nature Materials* **21**, 445-454 (2022).
32. Fan, X., *et al.* Non-flammable electrolyte enables Li-metal batteries with aggressive cathode chemistries. *Nature Nanotechnology* **13**, 715-722 (2018).
33. Wang, C., *et al.* Li3PO4-Enriched SEI on Graphite Anode Boosts Li+ De-Solvation Enabling Fast-Charging and Low-Temperature Lithium-Ion Batteries. *Angewandte Chemie* **136**, e202402301 (2024).
34. Li, Y., *et al.* Interfacial engineering to achieve an energy density of over 200 Wh kg− 1 in sodium batteries. *Nature Energy* **7**, 511-519 (2022).

## Acknowledgments

This work was supported by the National Key R&D Program of China (2022YFB2402500 and 2023YFC3304802), National Natural Science Foundation of China (52122214, 52072403 and U2268202), Jiangsu Province Carbon Peak and Neutrality Innovation Program (Industry tackling on prospect and key technology BE2022002-5), and financial support is greatly acknowledged from the Zhejiang Lab's International Talent Fund for Young Professionals.

## Author contributions

Y.Q.L., H. Z., S. Z., Z.H.L, and H.L. conceived the project and designed the experiments. Y.Q.L., Z.C., and Y.L. carried out electrochemical experiments and materials characterization, H.Z., X.F.G., and Y.Q.L. conducted the design and implementation of machine learning model algorithms. Y.Q.L., H. Z., X.F.G., S. Z., Z.H.L, and H.L. wrote the paper. All the authors participated in analysis of the experimental data and discussions of the results and in preparing the paper.

## Competing interests

The authors declare no competing interests.

## Data & Code availability

The data that support the findings of this study are available within this article and its Supplementary Information. Additionally, the large electrochemical dataset collected under simulated real-world cyclic protocols, the post-failure interfacial chemistry data, and the related code developed in this work will be made available following the official publication of the article.

# Supplementary Information

## Chemistry-aware battery degradation prediction under simulated real-world cyclic protocols

Yuqi Li, *et al.*

Corresponding author: shun.zheng@microsoft.com; zihenglu@microsoft.com; hli@iphy.ac.cn

**Table of contents:**

## Materials and Methods

### Electrochemical testing and interface chemical analysis

The dataset used in this study was collected from Samsung ICR 18650-22P cylindrical batteries with a rated capacity of 2.2Ah. The charge and discharge tests of the cylindrical batteries were performed on the Neware battery test system (CT-4008), with the batteries being charged and discharged between 2.75 V and 4.2 V. The cells were placed in an environmental chamber at different temperatures.

The morphologies of the electrode surface were examined using a Hitachi-S4800 scanning electron microscope (SEM). Digital photographs were captured by iPhone X. The surface chemical state of the anode was determined by X-ray photoelectron spectroscopy (XPS) using the Thermo Scientific ESCALAB 250 Xi spectrometer with an Mg/Al Kα X-ray source. All spectra were calibrated with the C 1s photoemission peak at 284.8 eV to correct for the charging effect. For the ex-situ measurements, the electrode disassembled from the cylindrical battery in an argon-filled glove box was washed in dimethyl carbonate (DMC) repeatedly, followed by a drying process in an argon-filled transfer tube.

### Random Operating Condition Generation

In our investigation, we focused on highway driving scenarios as outlined by the U.S. Environmental Protection Agency's (EPA) US06 aggressive driving schedule[1] with high acceleration. This schedule was transformed into battery power profiles through a vehicle dynamics model[2]. To align with the maximum output capacity of our testing device, we adjusted the power values, ensuring they did not exceed 16W. Additionally, we refined the cycling process by subsampling zero power instances in the original profile, effectively shortening idle periods in our discharging protocols.

The dynamics of power consumption, analogous to vehicle speed, can be considered a stochastic process. Here, the vehicle's speed at any given moment is influenced by its preceding conditions, including driver behavior, traffic condition and the vehicle's state. Instead of directly modeling this complex stochastic process, which becomes growingly complex as the number of time steps increase, we propose a first-order Markov assumption for simplification. We introduce a hidden variable to summarize past conditions, encapsulating all influence factors such as the driver's habits, the vehicle's current state, and road conditions. This simplification streamlines training and inference to linear time, while accurately captures the stochastic nature of the speed variation.

Adopting this first-order Markov perspective, we utilize a Hidden Markov Model (HMM)[3] to represent the dynamics of the unobservable state, informed by observed driving patterns. Given the continuous nature of vehicle speed and its usual confinement within certain ranges due to speed limits, we incorporate Gaussian emissions within the HMM framework. Here, the speed observed at time t, represented by $x_t$, follows a Gaussian distribution characterized by a mean $\mu_{ht}$ and variance $\sigma^2_{ht}$, where $h_t$ denotes the hidden state at time t. The transition to the next state, $h_{t+1}$, is governed by a probability matrix $A$, with $A_{ij}$ indicating the likelihood of transitioning from state $i$ to state $j$. The optimization of this model involves adjusting the transition matrix and the parameters of the Gaussian distributions for each hidden state by maximizing the likelihood. Post-optimization, we generate novel, random driving protocols by sampling an initial hidden state and subsequently producing observations and transitions in an autoregressive manner.

This approach not only ensures compliance with hardware limitations but also accelerates the testing process without compromising the integrity of the driving scenarios being simulated. By leveraging a hidden variable to encapsulate complex dependencies, our method offers a scalable and efficient means of modeling and predicting vehicle power consumption under varied driving conditions.

**Enhanced Batch Battery Testing Scheduling Framework**

Current battery cyclers are adept at conducting efficient testing of specific cycling protocols on multiple cells simultaneously. This approach is critical for minimizing variance caused by the inherent differences between individual cells, thereby ensuring robust statistical analysis. However, this methodology necessitates manual intervention upon the completion of each protocol. This requirement becomes problematic when dealing with random protocols that require multiple specification files to reach the end of life, as the varied testing durations for different cells make manual intervention impractical. Such delays can inadvertently introduce periods of inactivity for the batteries, adversely affecting their degradation paths.

To address these challenges, we have developed a comprehensive framework that facilitates flexible scheduling of batch testing. Our system is compatible with battery cyclers that can both execute any specified cycling protocol and report the current cycling status. It operates by launching independent threads for each battery testing channel to initiate and monitor the cycling tests. These monitoring threads continuously poll the cycler's status to log the number of cycles completed and the specifications used, ensuring progress is recorded for potential recovery in the event of a power outage. Upon completing a test for a given specification file, the monitoring thread submits an execution summary to a central master thread before being terminated. The master thread then generates a new monitoring thread to proceed with subsequent tests, if available. Through this mechanism, our framework introduces a versatile scheduling layer capable of accommodating various cyclers, streamlining the process and significantly reducing the need for manual intervention.

**Automated Polynomial-Scale Feature Space Generation**

Current approaches to designing feature spaces involve manual intervention and rely heavily on domain expertise, which presents two major challenges that limit their applicability to randomized protocols. First, these methods assume uniformity across adjacent cycles, which leads to similar patterns that ensure algorithmic insensitivity to cycle selection and generalizability across battery types. However, under randomized protocols, cycle discrepancies lead to significant variation in electrical signals, undermining direct comparisons and generalization capabilities. Second, the assumption of smooth input signals throughout the cycling process is crucial for accurately reflecting early-cycle degradation. Random protocols introduce statistical noise, significantly diminishing the predictive accuracy of machine learning models.

To overcome these limitations, we introduce a novel methodology that autonomously generates a feature space on a polynomial scale. This process involves several levels of composition of feature extractors. We first partitioned the cycles into K groups and converted each cycle's time-series data—such as charging and discharging voltages—into fixed-length, capacity-indexed curves. These curves are then segmented into D parts, with statistical operators applied to each segment to derive local features. We mainly employed operators like the mean, min, max, variance, skewness and kurtosis, as these statistical measures highly determine the moment generating function of the random variable. Finally, by aggregating these features with specific function compositions, we obtained a feature space that expands polynomially with K, D, and the number of operators used.

Integrating all cycles into the feature construction process provides a holistic view of the degradation process, essential for accurately modeling under randomized protocols. Segmenting cycles into distinct groups enables the model to identify and extract features that signify specific aging phenomena, such as the early activation stage or potential accelerated degradation during battery testing.

Dividing the electrical signals into smaller segments diminishes the variability introduced by random protocols. These localized segments, characterized by reduced workload transitions and smoother patterns, are more conducive to effective feature extraction. Nonetheless, there's a limit to how finely we can divide these signals without unnecessarily expanding the feature space. Over-segmentation leads to computational challenges and a heightened risk of overfitting due to the increased dimensionality. Furthermore, these segments offer insights into various phases of the charging and discharging cycles, enabling the model to distinguish and learn from the nuances between rapid voltage drops and the stable plateau phases.

**Data-driven Degradation Learning Pipeline**

For all predictive tasks in this study, we adopt the random forest model to balance the accuracy and efficiency. Random forests consist of a series of decision trees which independently fit on a random subset of data samples using a random feature subset and ensemble for accurate predictions. Considering the diverse degradation patterns brought by the random protocols, random forest is well-suited for modeling battery aging tasks in this context, both in terms of accuracy and interpretability. The different base decision trees may capture different patterns for more effective learning, and the tree structure naturally allows users to interpret the predictions made. Such a design allows the model to adapt without modifications to different downstream battery aging modeling tasks, demonstrating outstanding performance in the tasks detailed below.

In this study, we investigate modeling various battery aging objectives within the proposed feature space. For the task of early prediction of battery lifetime, we estimate the number of spent cycles for each battery to decay to 80% of its nominal capacity. Specifically, due to the wide temperature range studied in this paper, even batteries of the same model exhibit significant differences in discharge capacity during early cycles at different temperatures, greatly affecting cycle life calculation (batteries at low temperatures can drop below 80% of the suggested nominal capacity in early cycles). Therefore, we take the average of the discharge capacities of the first 5 cycles for all batteries at each temperature as the nominal capacity of this battery type at that temperature. Based on this definition, we calculate the cycle life of these batteries independently under each temperature. We employed a random forest regression model to recover the battery's lifespan given only the first 50 cycles. Supplementary Table 1 confirms the predictive capability of this automatic learning pipeline in forecasting battery degradation. In the knee point prediction task[4], we calculated the slope of discharge capacity decrease for each battery at every 50-cycle interval. We determined the existence of a knee point in the discharge capacity degradation curve by comparing the maximum slope of these intervals to a pre-determined threshold (0.0005 in this study). We utilized the proposed feature space with a random forest classifier to predict the existence of knee point. For intuitive illustration of the task setting, Supplementary Fig. 9a illustrates a cell with a significant knee point around cycle 50, while Supplementary Fig. 9b shows a battery that degrades almost uniformly, lacking a distinct knee point. Supplementary Fig. 9c presents the distribution of the max degrade change across all batteries. For the interface chemistry prediction task, we collected five sample points from each disassembled battery at the end of its lifecycle. XPS measurements were then performed on these samples to analyze various elements, both before and after etching. This process yielded comprehensive data from 56 batteries, with most batteries providing XPS measurements from all five samples. In total, 552 sets of raw XPS data were gathered, encompassing both pre-etching and post-etching measurements. One battery had incomplete data, resulting in slightly fewer samples. We used K-means clustering[5] to categorize all 56 batteries into 8 groups (determined intuitively with the elbow method) based on the proportions of the elements 'Li1s', 'C1s', 'O1s', 'F1s', 'P2p', 'Ni3p1', 'Co3p1', and 'Mn3p' , as detailed in Supplementary Table. 5. Notably, two of the detected battery groups, Group 4 and Group 5, contained only one battery, which consistently failed to cluster with other battery groups due to highly unsimilar XPS patterns. Therefore, we excluded these two batteries in our investigation and focused on the six XPS pattern groups formed by the 54 remaining batteries. We fed the proposed automatically generated polynomial-scale feature space to a random forest classifier to predict which pattern group this battery belongs to. Surprisingly, this feature space based entirely on early electrical signals successfully captures the patterns of the interface chemistry. This confirms that the electrical signals as a non-intrusive data source can reflect the underlying chemistry process using the proposed learning pipeline.

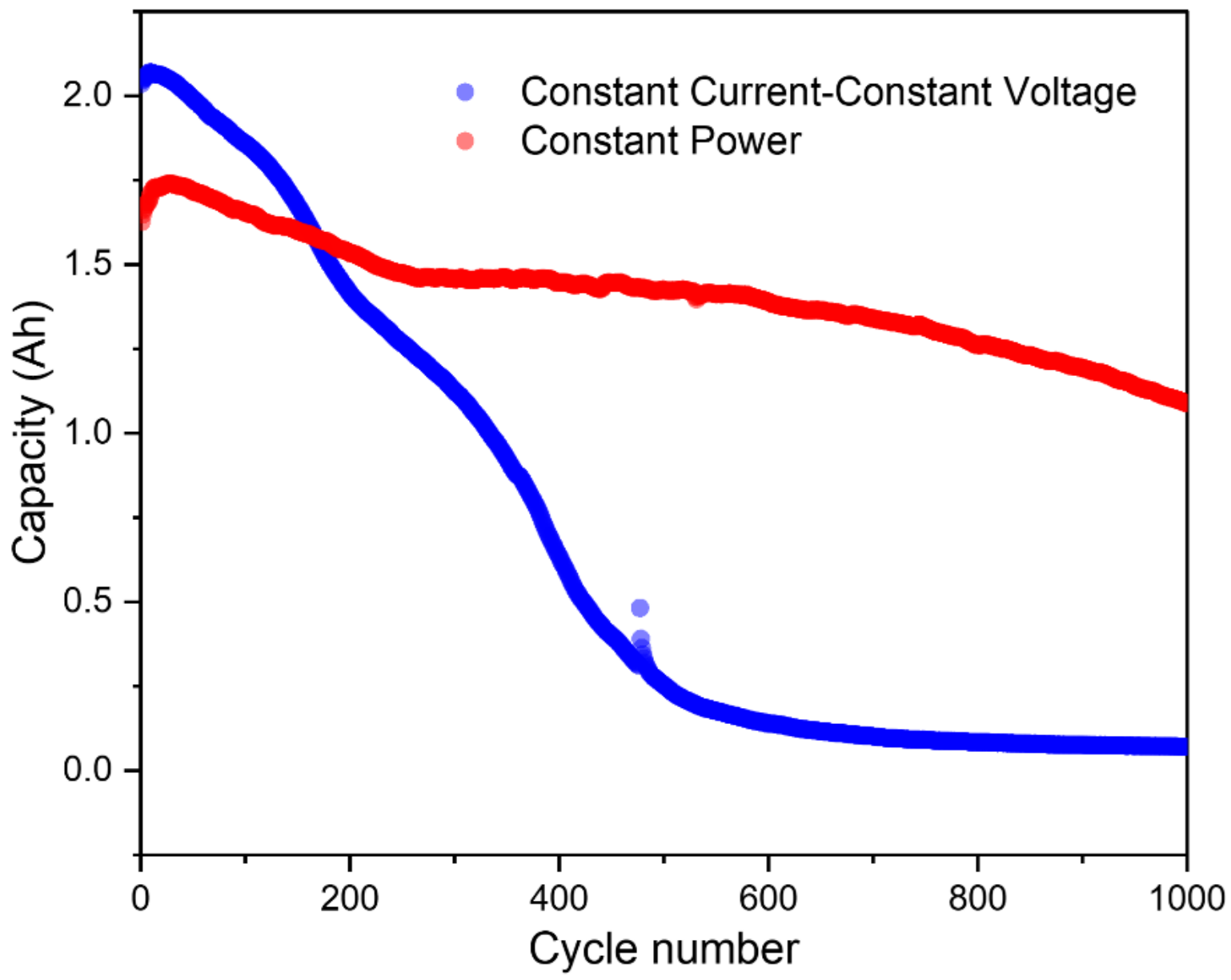

**Supplementary Fig. 1.**

Cycling tests of 18650 power-type lithium-ion batteries under constant current-constant voltage (CC-CV) and constant power (CP) conditions.

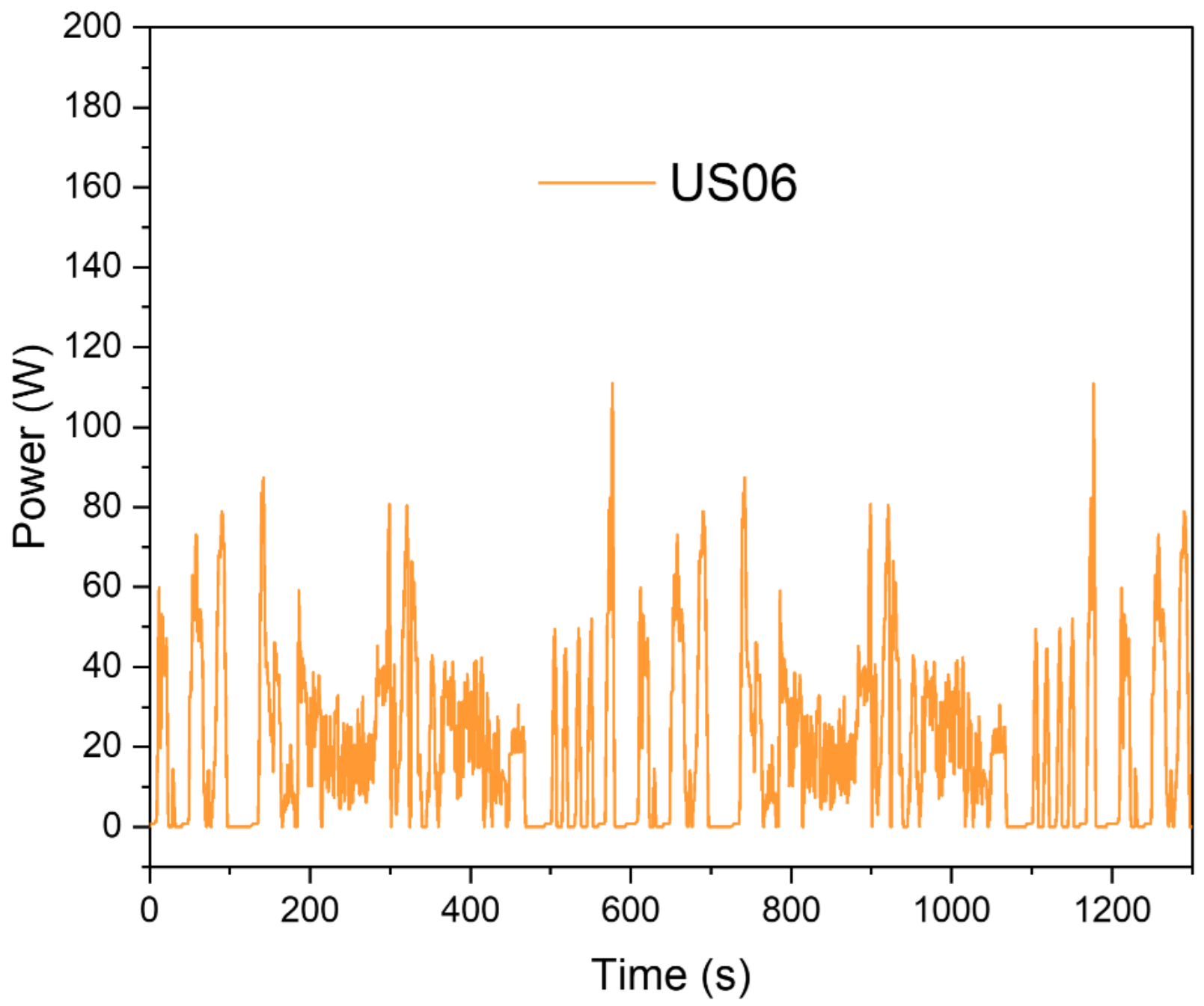

**Supplementary Fig. 2.**

The highway driving profiles of the US06 high acceleration aggressive driving based on the parameters of 18650 power-type lithium-ion batteries.

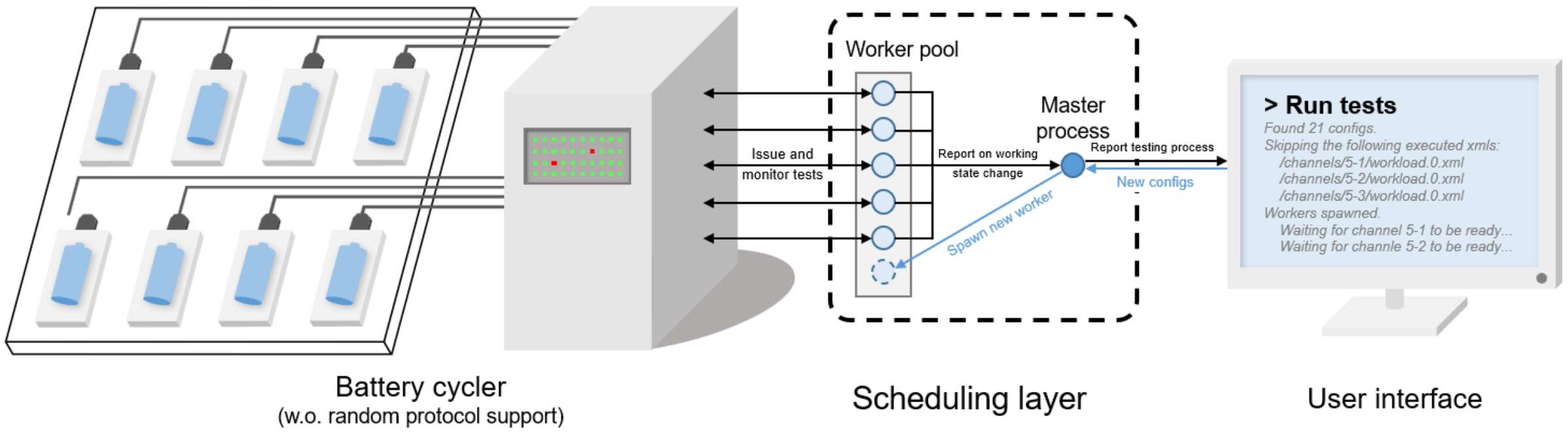

**Supplementary Fig. 3.**

Multi-tasking controllers for parallel battery testing channels, which can automate the entire data collection process that starts, monitors, and stops battery tests in each test channel with different stochastic discharge protocols.

Despite the capability to generate simulated real-world cyclic protocols, a challenge remains on the battery testing level due to the inability to manually control these complex steps. A sufficiently large configuration file is essential to comprehensively cover the entire lifespan of a battery under real-world cycling protocols. However, most battery cyclers on the market typically lack the capability to handle such extensive configurations. If the configuration is split into multiple files, manual intervention becomes necessary to manage the testing process, which not only consumes substantial human resources for continuous monitoring but also inevitably introduces undesired rest periods. This interruption can potentially skew the test results, making it difficult to obtain accurate data for predicting battery performance in real-world conditions. This suggests a pressing need for the development of multitasking processors dedicated to the automatic control of battery operations. Therefore, we developed a batch-testing system that supports flexible cycling protocols on existing battery testers (Fig.1b and Supplementary Fig. 3). By utilizing a multi-tasking controller for parallel scheduling of battery testing channels, we automated the entire data collection process that starts, monitors, and stops battery tests in each test channel with different stochastic discharge protocols, significantly enhancing testing efficiency.

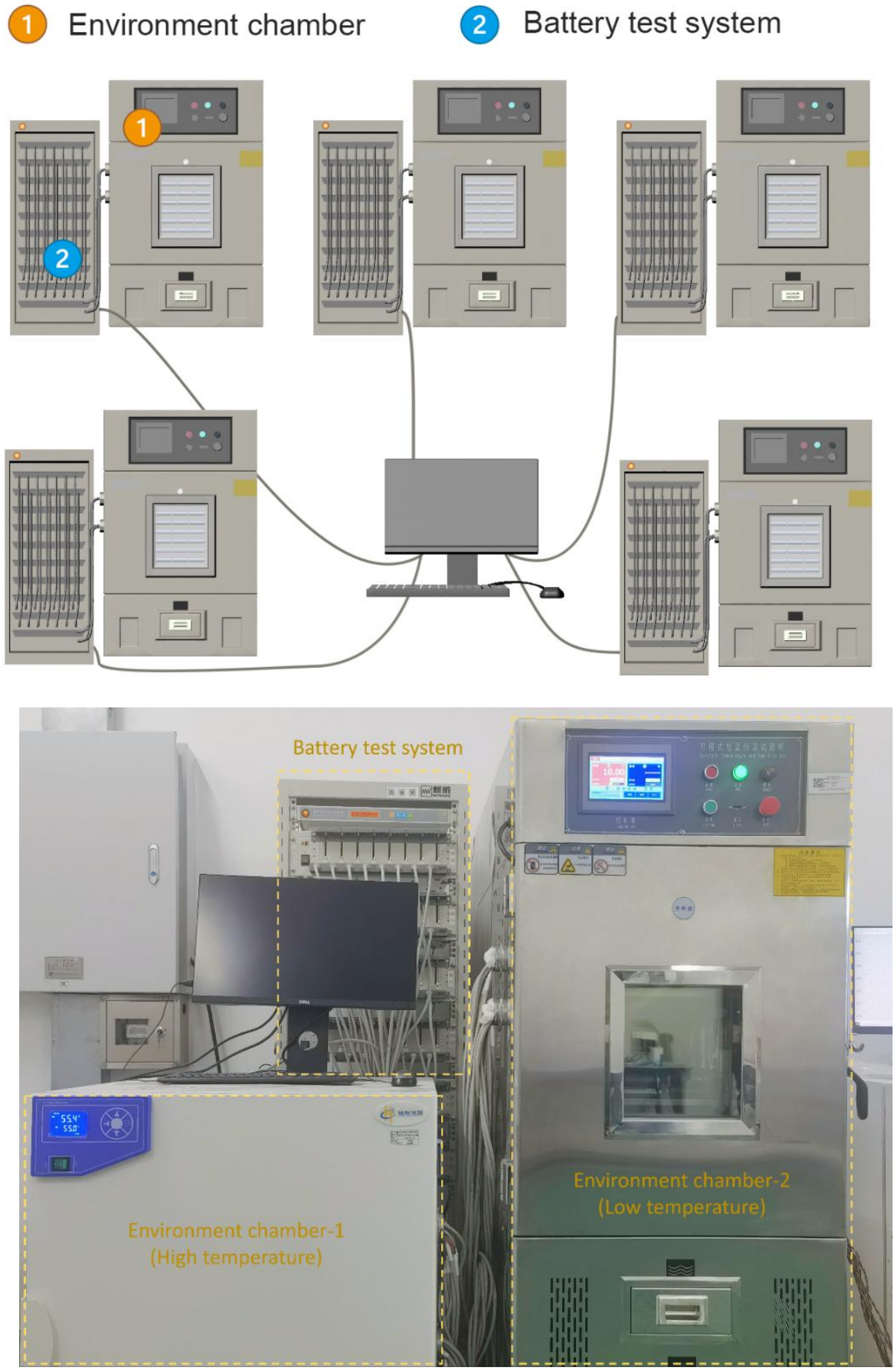

**Supplementary Fig. 4.**
The schematic connection of the battery test system, environment chambers, cells, and PC.

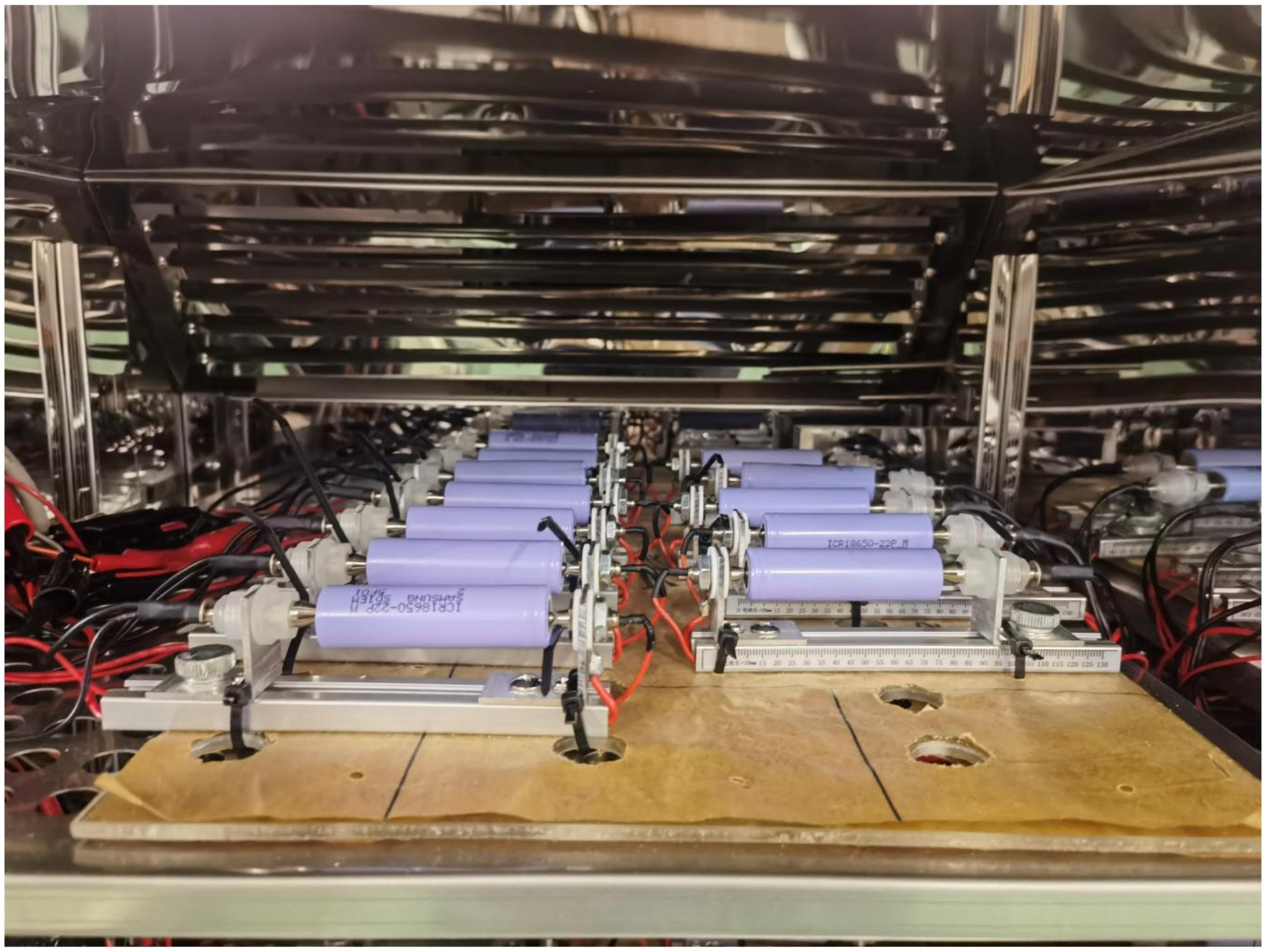

**Supplementary Fig. 5.**

Connection of the cylindrical cells inside the environmental oven. A four-wire holder is used for the batteries.

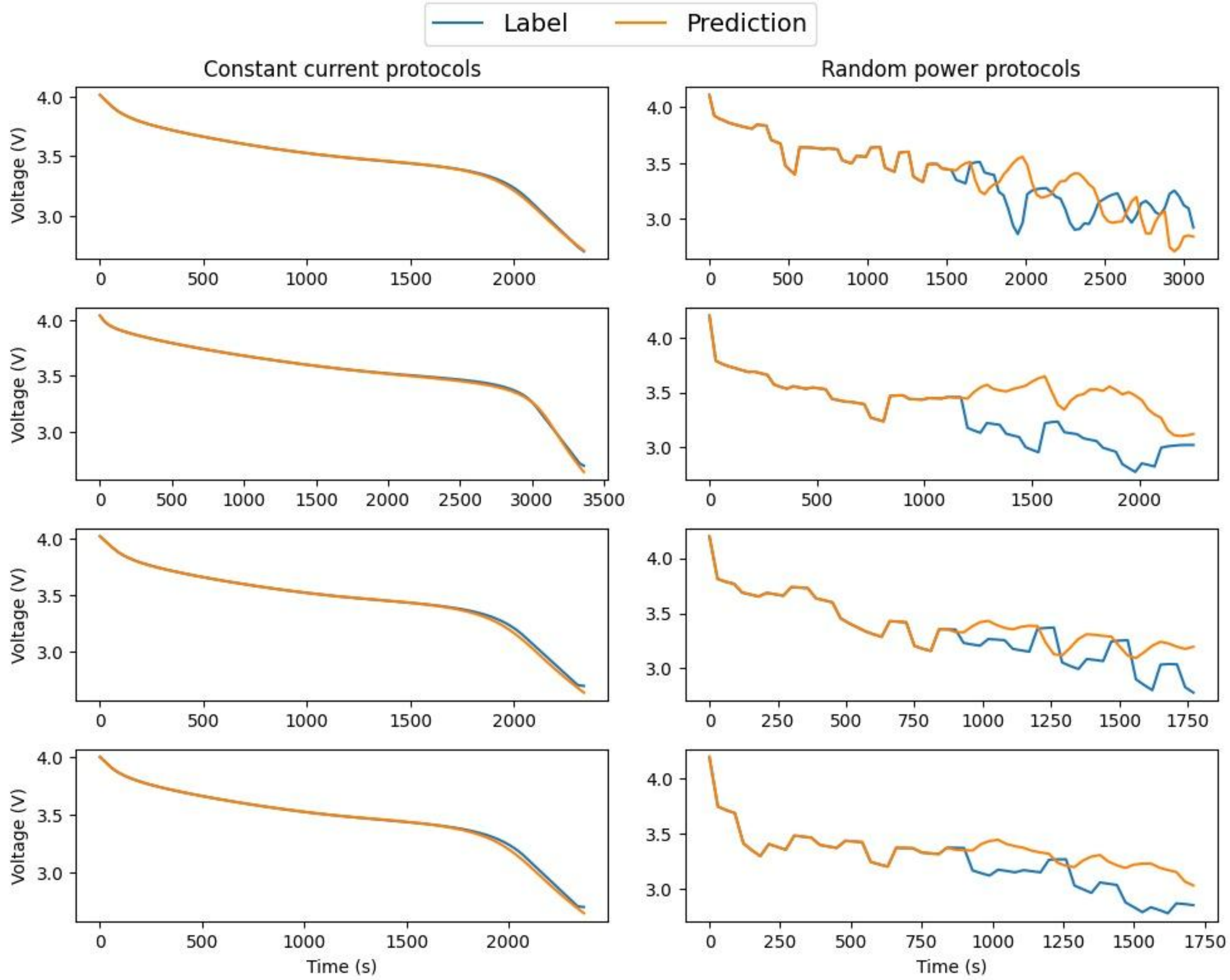

**Supplementary Fig. 6.**

Predicted electrochemical curves under constant and random discharge protocols. The learning approach involves predicting future changes in current and voltage based on past values, with predictions starting after 50% for both cases.

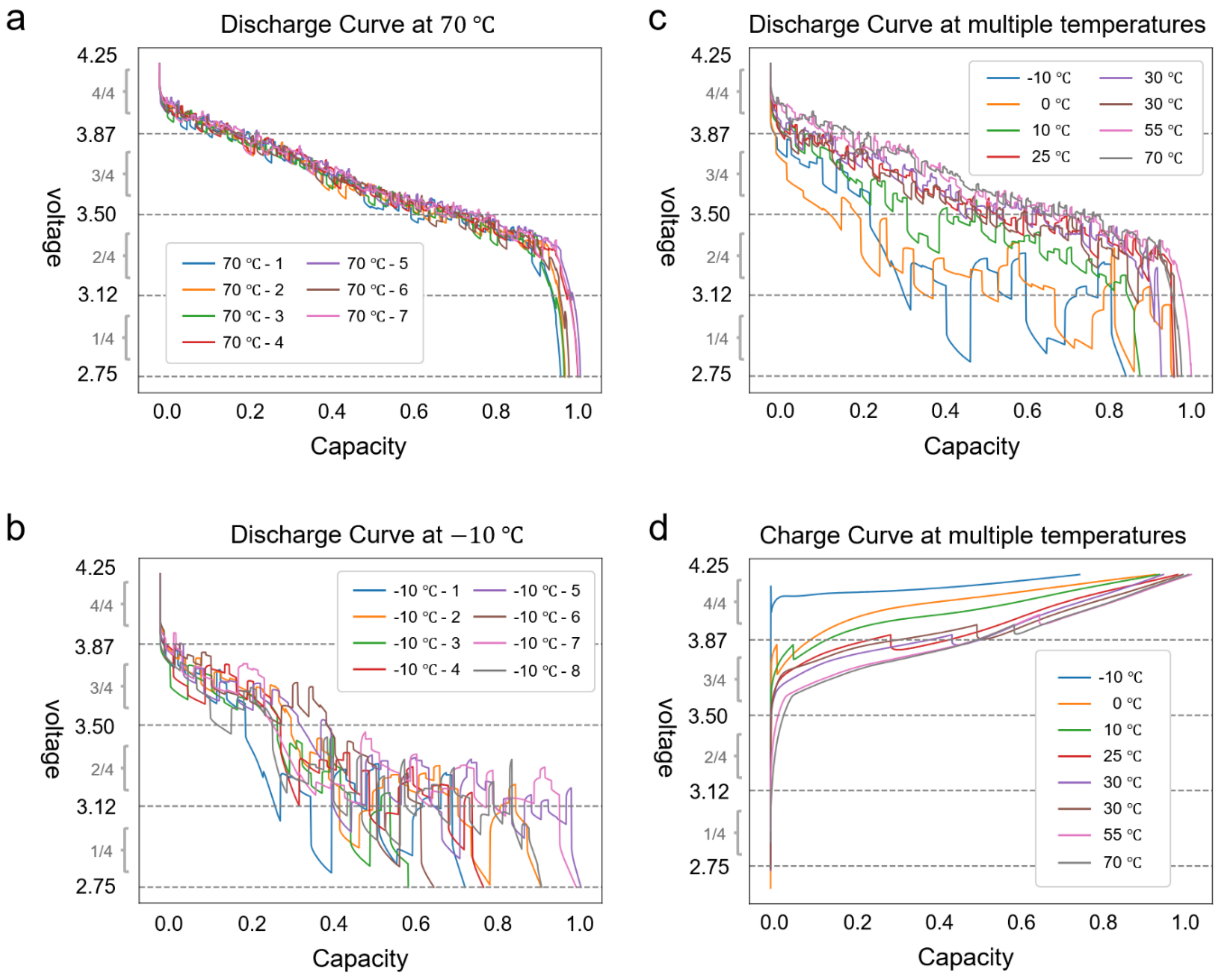

**Supplementary Fig. 7.**

Typical charge and discharge curves (second cycle) under random operating conditions.

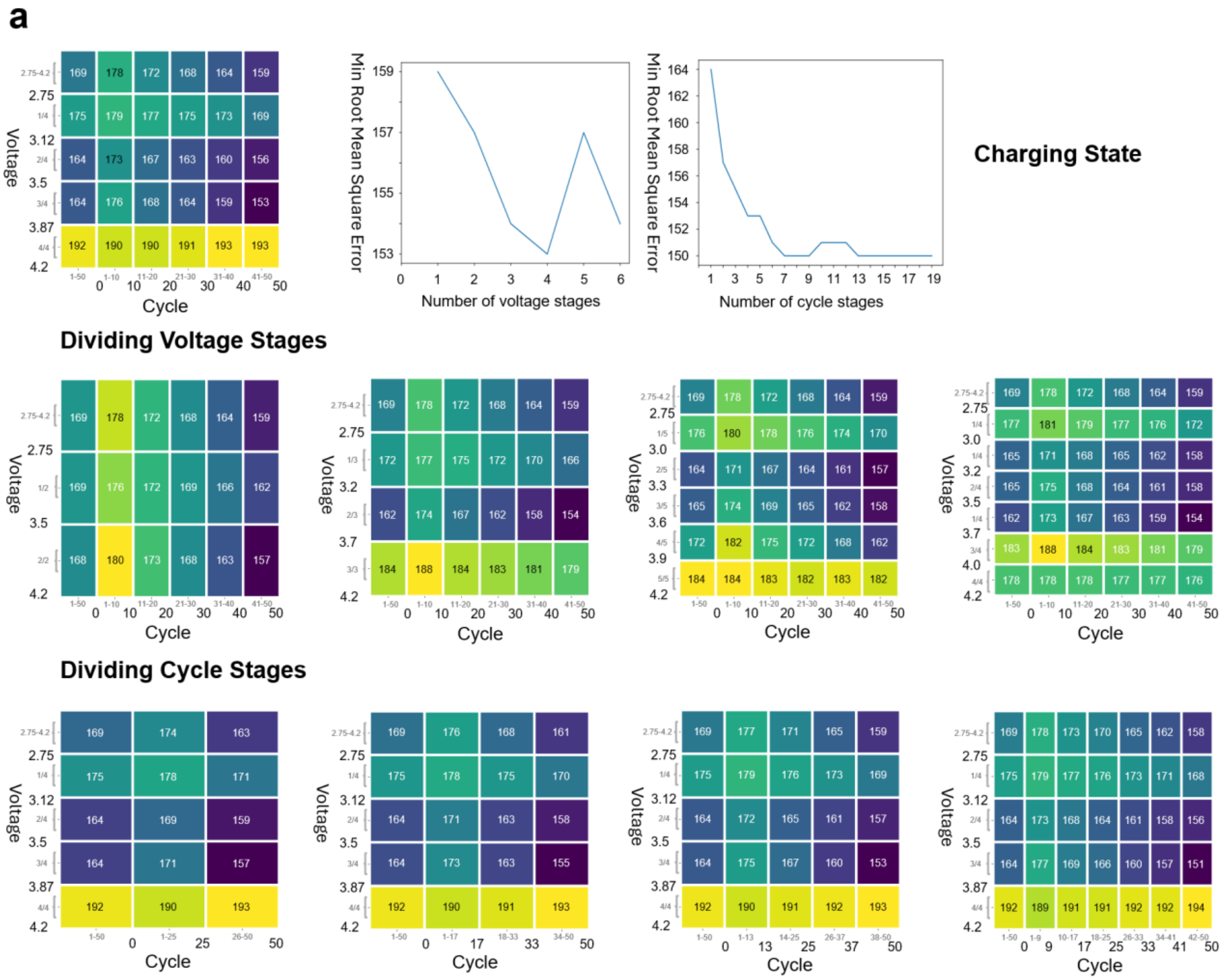
a
Charging State
Dividing Voltage Stages
Dividing Cycle Stages
Voltage
Cycle
Min Root Mean Square Error
Number of voltage stages
Number of cycle stages

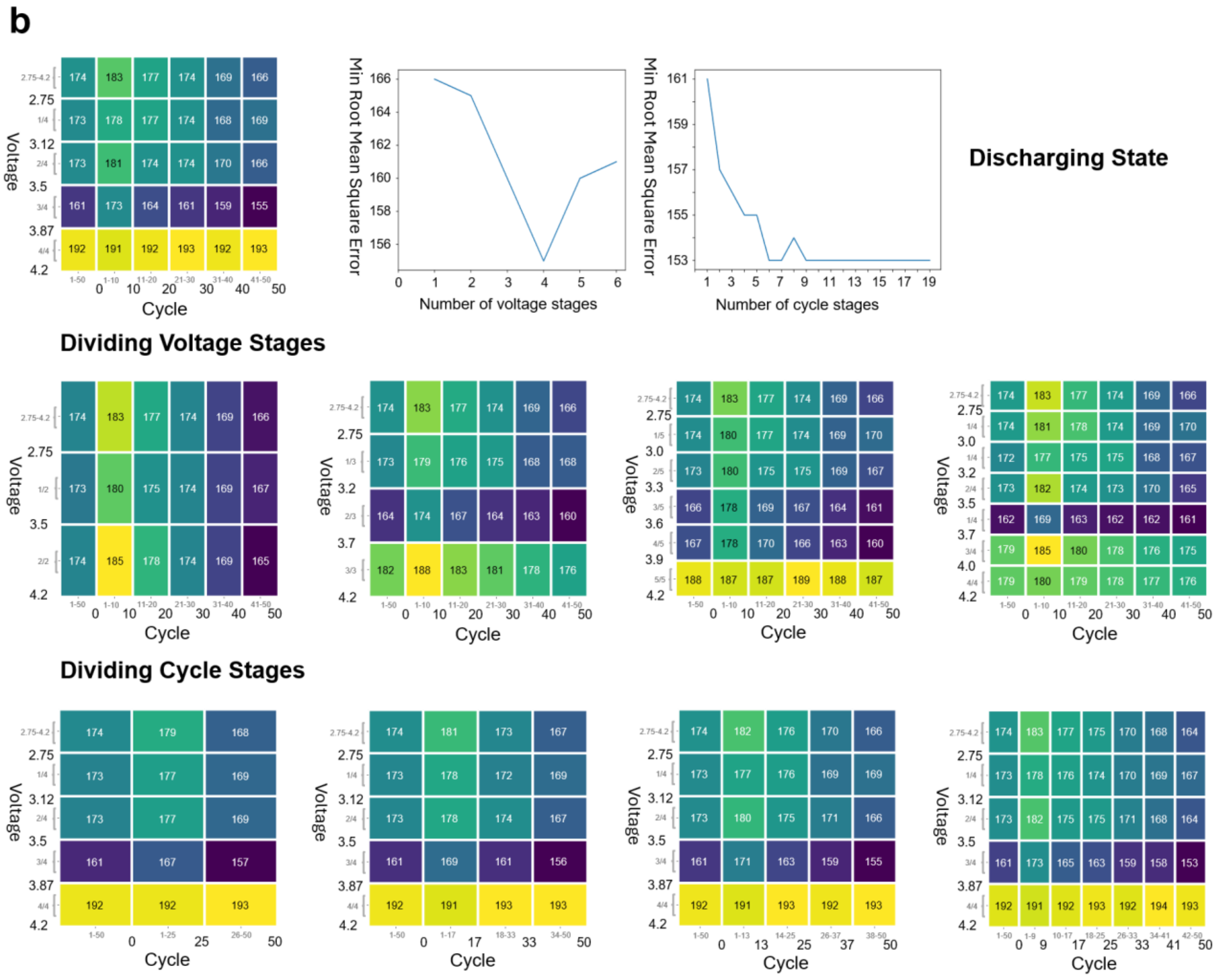

**Supplementary Fig. 8.**

Heatmap of the relationship between prediction errors and voltage segmentation and cycle segmentation in feature space construction. The splitting strategy—dividing the data into seven cycle groups and four signal segments—strikes a balance between complexity and performance. The feature format used is "identity(nanmean(Cycle(cycle index / cycle group))[nanmean($V_{Q_d/Q_c}$(intra-cycle index / intra-cycle group))])" a, A graph showing the relationship between battery cycles (x-axis) and capacity (y-axis) with significant knee points. b, A graph showing the relationship between battery cycles (x-axis) and capacity (y-axis) without significant knee points."Qd slope" refers to the slope of the curve between each pair of adjacent green anchor points. The "Qd slope change" is defined as the difference between the Qd slope of one segment and the preceding segment, specifically $Qd\ slope_{i+1} - Qd\ slope_i$. The "Qd slope change ratio" is calculated as the Qd slope change divided by the slope of the preceding segment, expressed as $\frac{Qd\ slope_{i+1} - Qd\ slope_i}{Qd\ slope_i}$. **c**, Distribution of the maximum rate of degradation change across all cells.

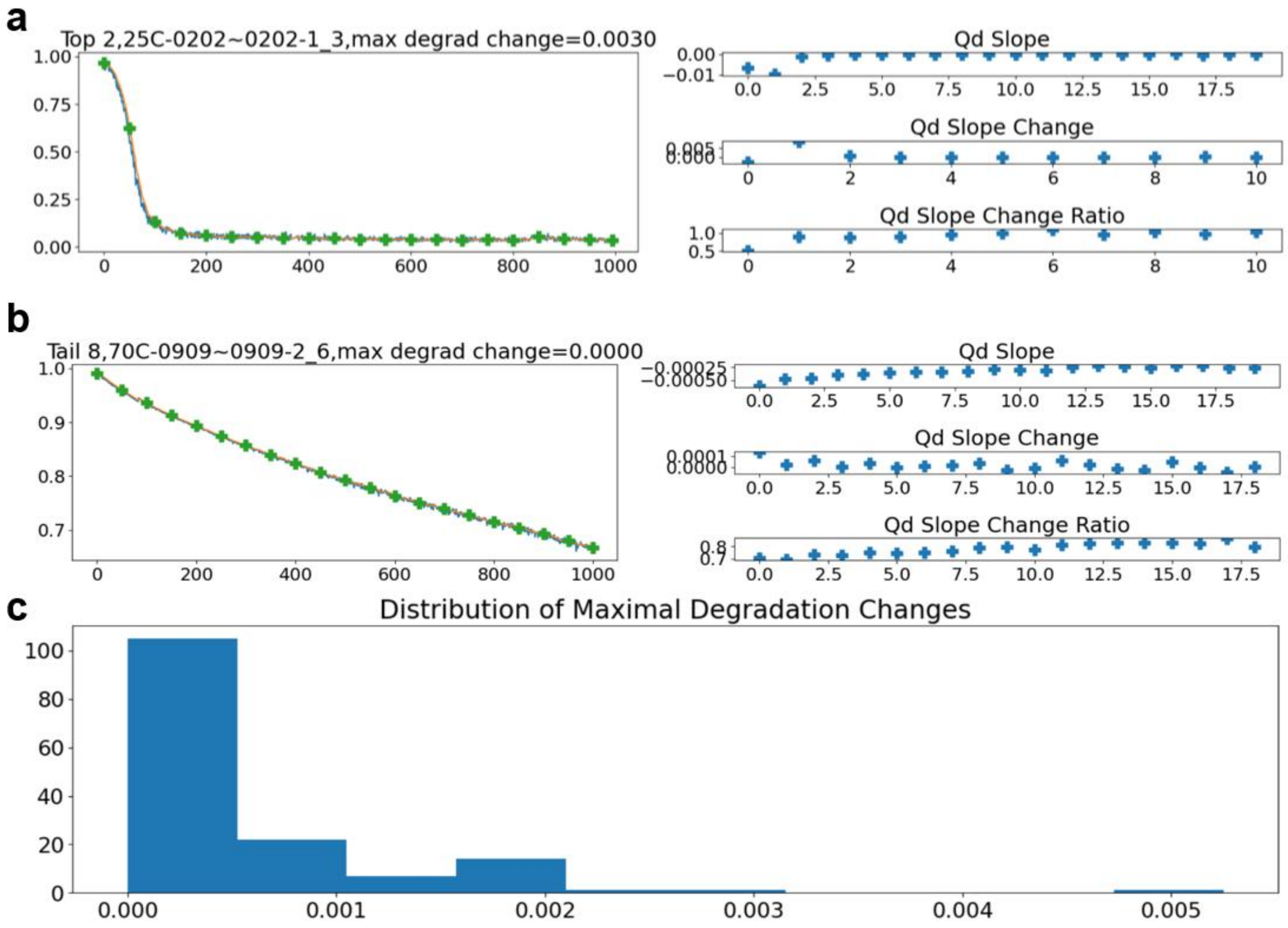

**Supplementary Fig. 9.**

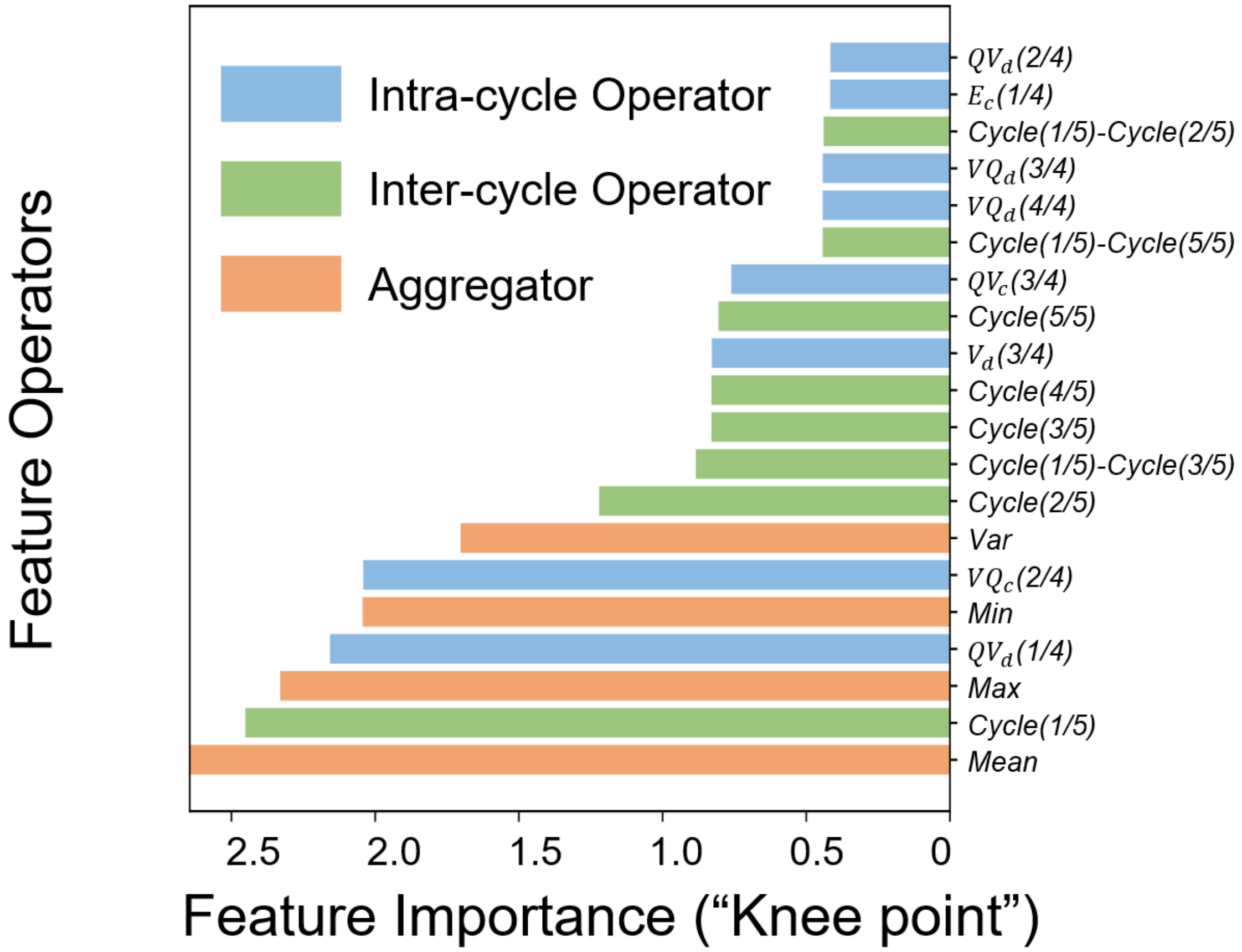

**Supplementary Fig. 10.**

The sub-operators of the most important features for the battery "Knee Point" prediction (based on their frequency of selection in random forest model, represented by "count") inputted into the machine learning model.

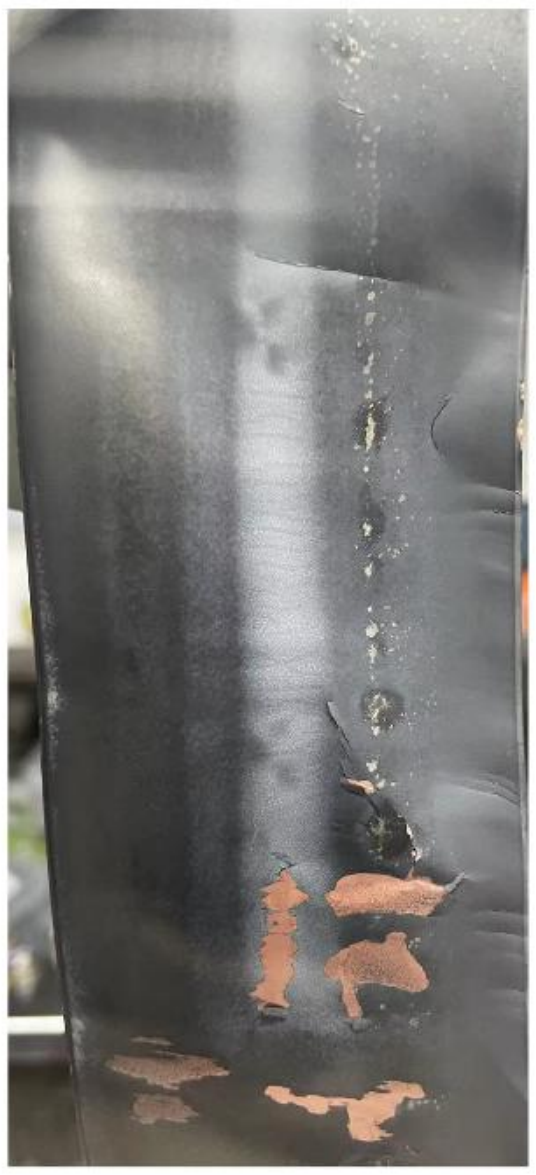

**Supplementary Fig. 11.**

Optical images of the disassembled graphite anode after 1000 cycles.

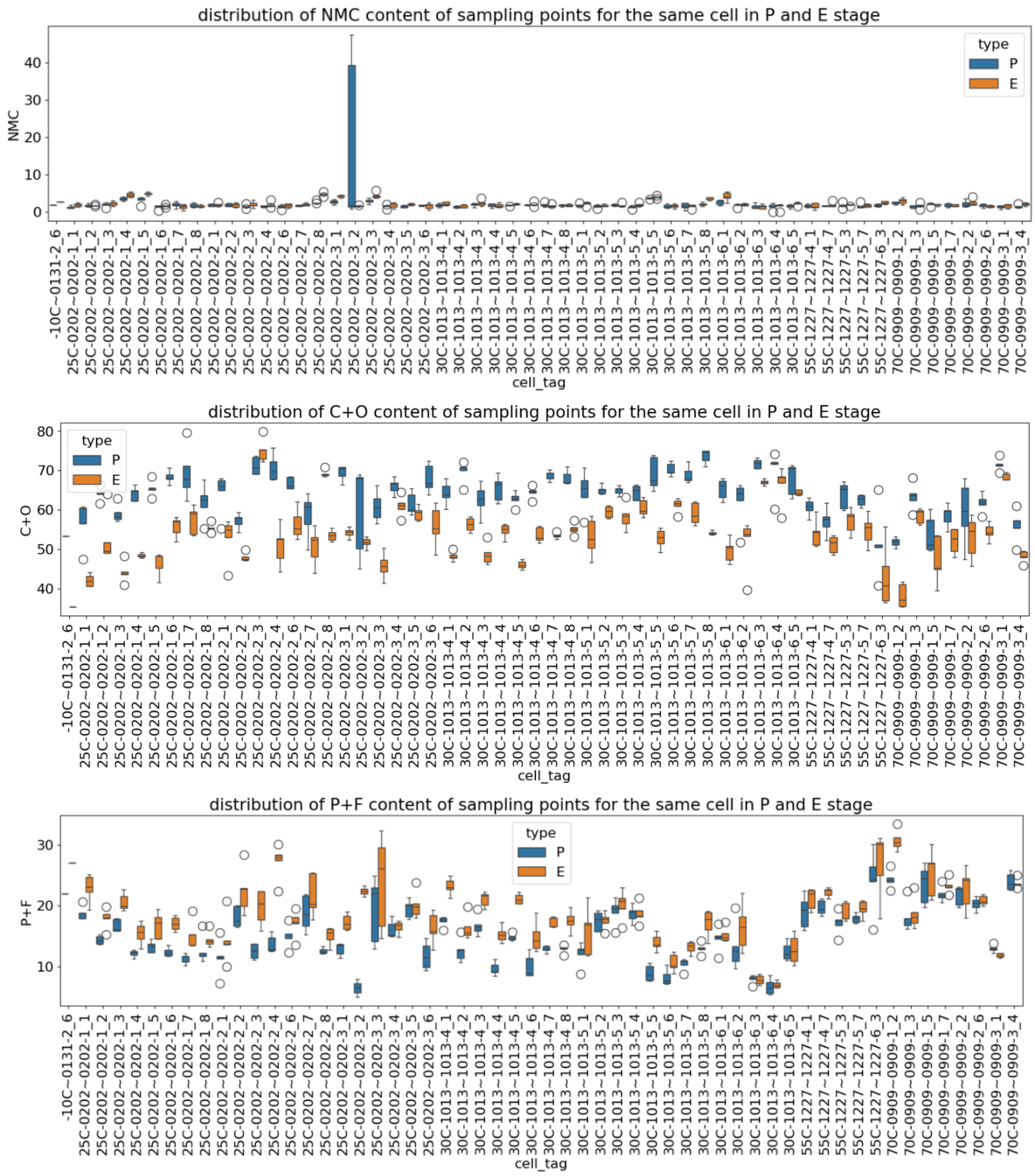

**Supplementary Fig. 12.**

The interface chemistry information of all failed batteries (P: Before Etching; E: After Etching; NMC=Ni+Mn+Co).

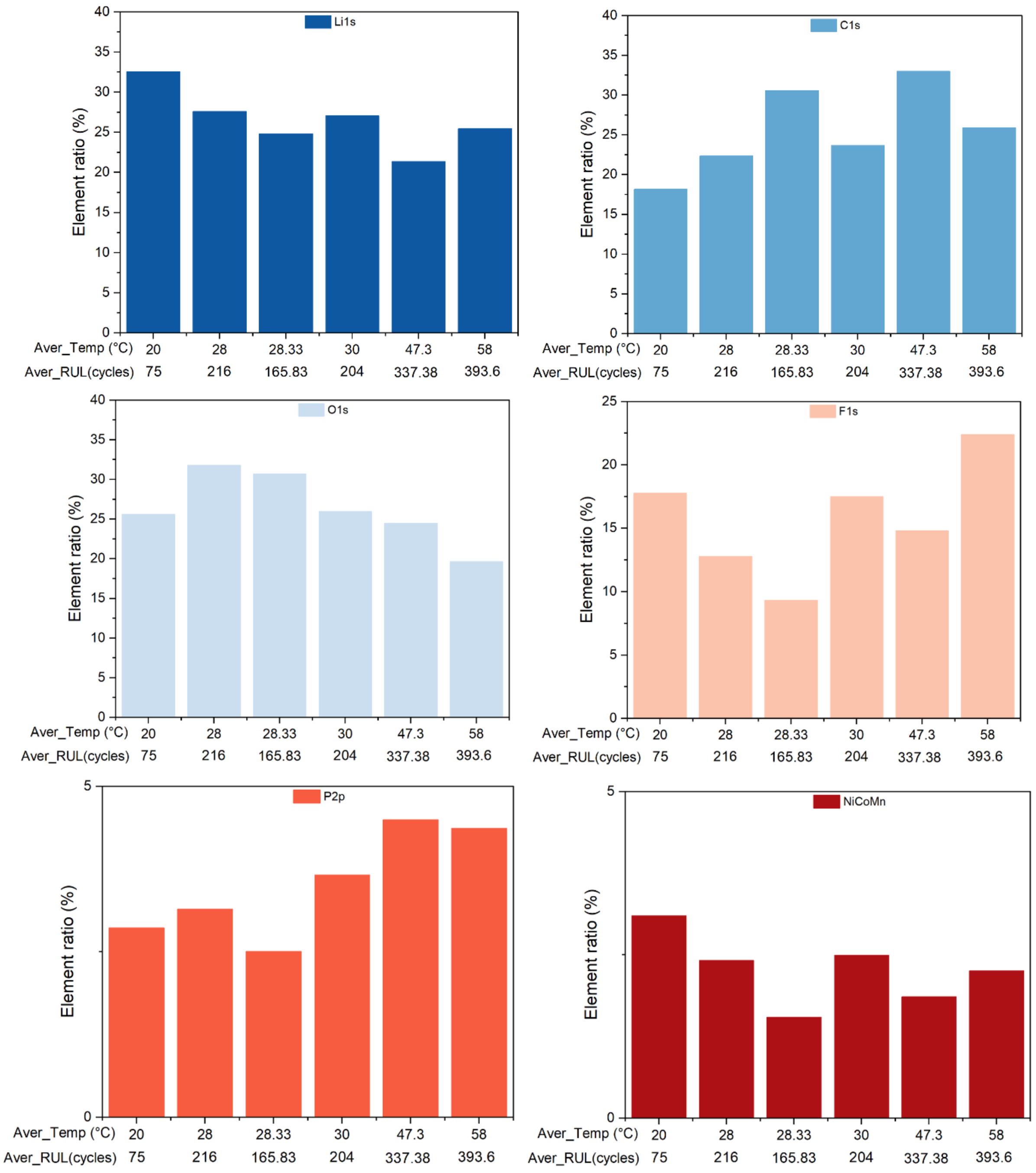

**Supplementary Fig. 13.**

Six types of XPS patterns generated by clustering, along with their corresponding average elemental ratios, average operating temperature (Aver_Temp), and average remaining useful life (Aver_RUL) of the battery.

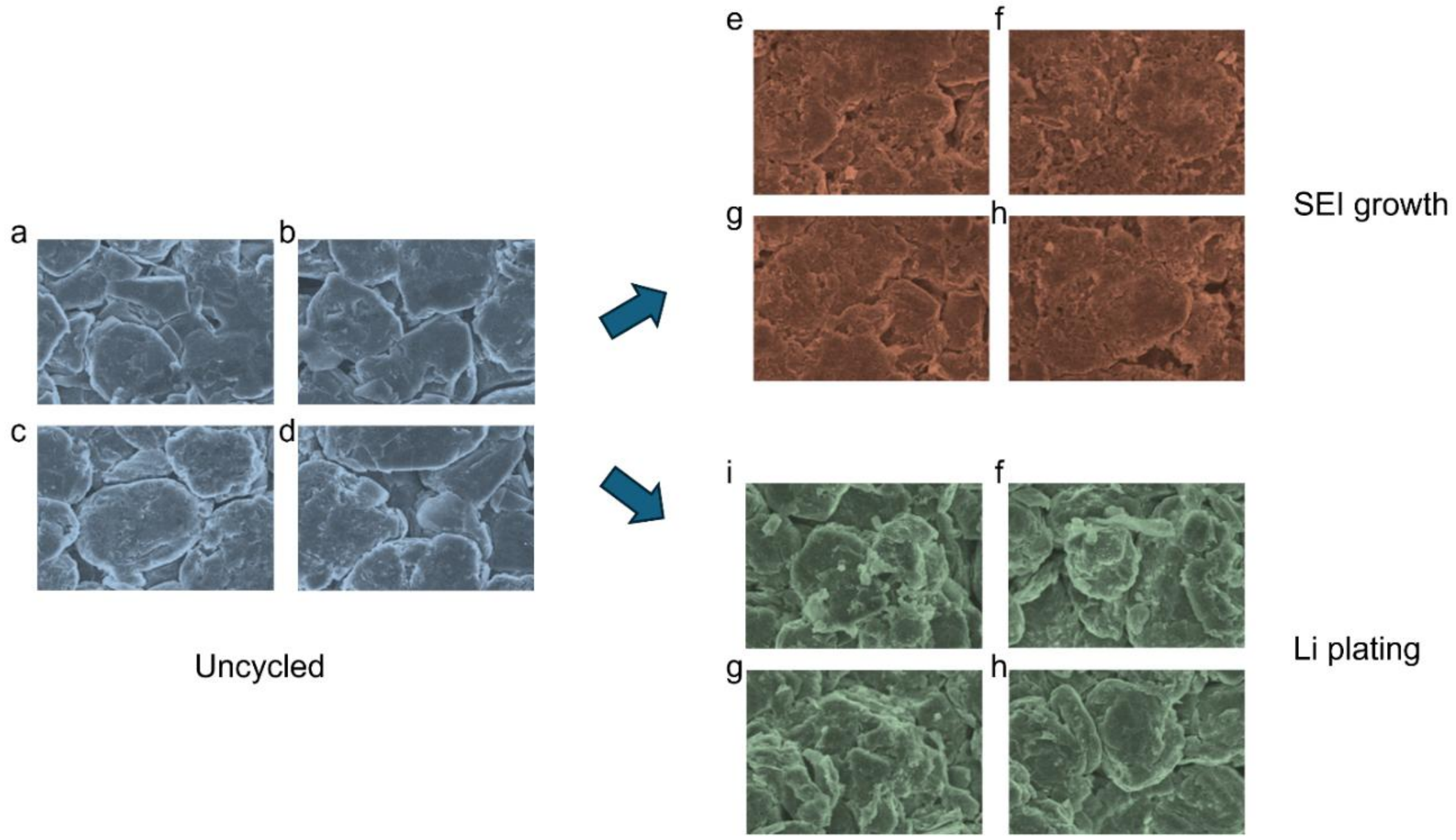

**Supplementary Fig. 14.**

SEM images of the graphite anode side under uncycled conditions, and with SEI growth (pattern 6) and Li plating (pattern 1).

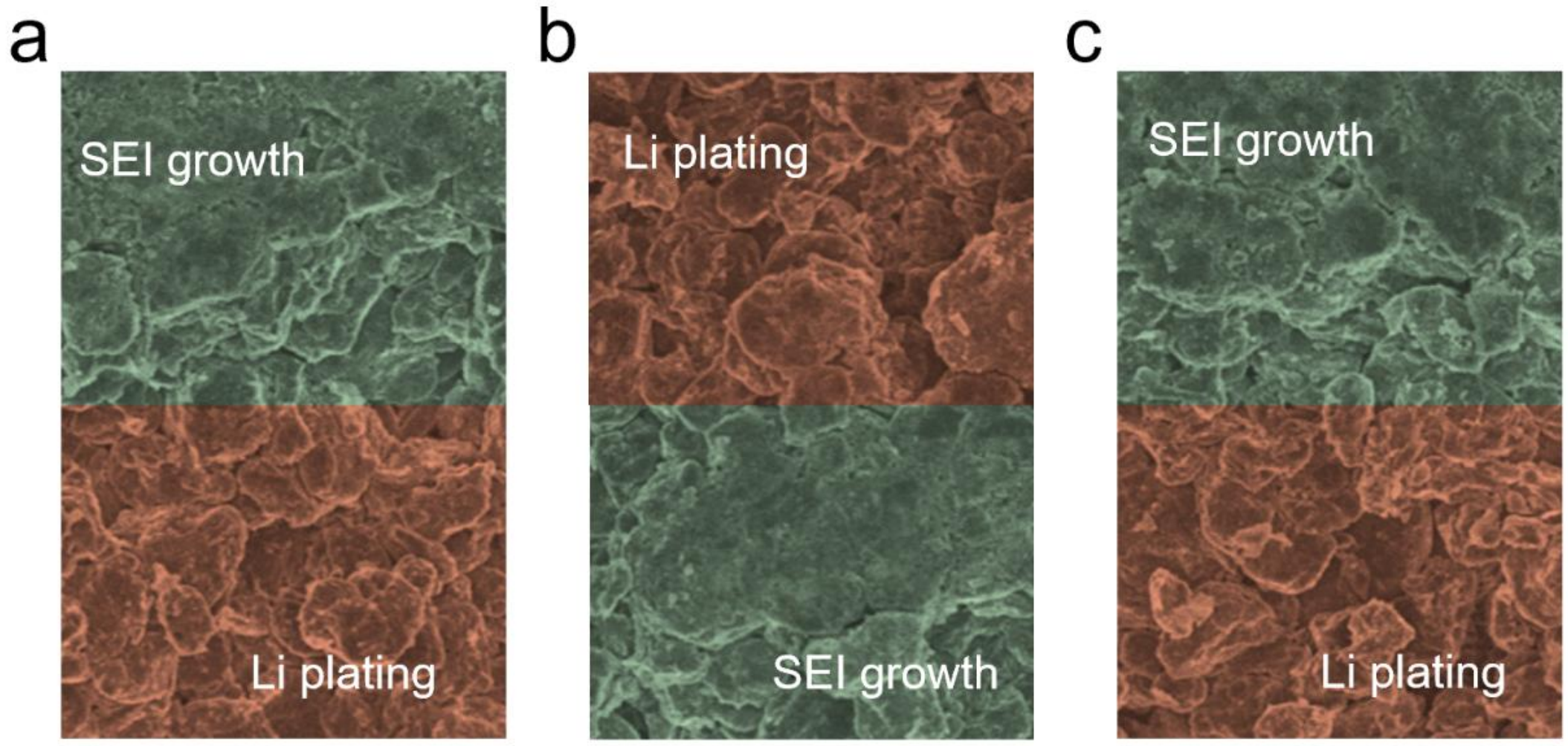

**Supplementary Fig. 15.**

SEM images of the graphite anode side (related to the patterns 2-4) showing hybrid failure mechanisms of Li plating and SEI growth.

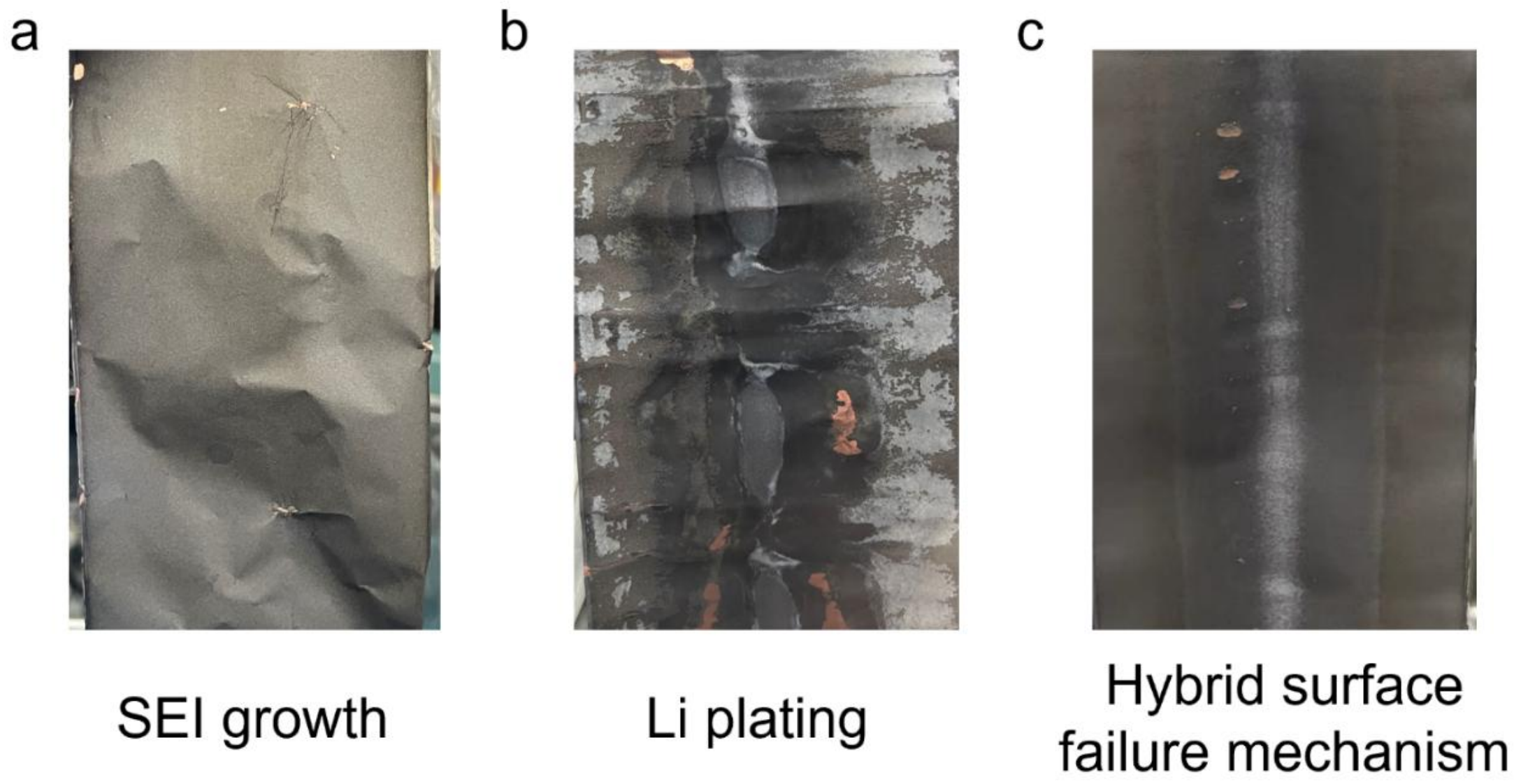

**Supplementary Fig. 16.**

Optical images of the graphite anode side under various conditions, showing SEI growth, Li plating, and hybrid failure mechanisms.

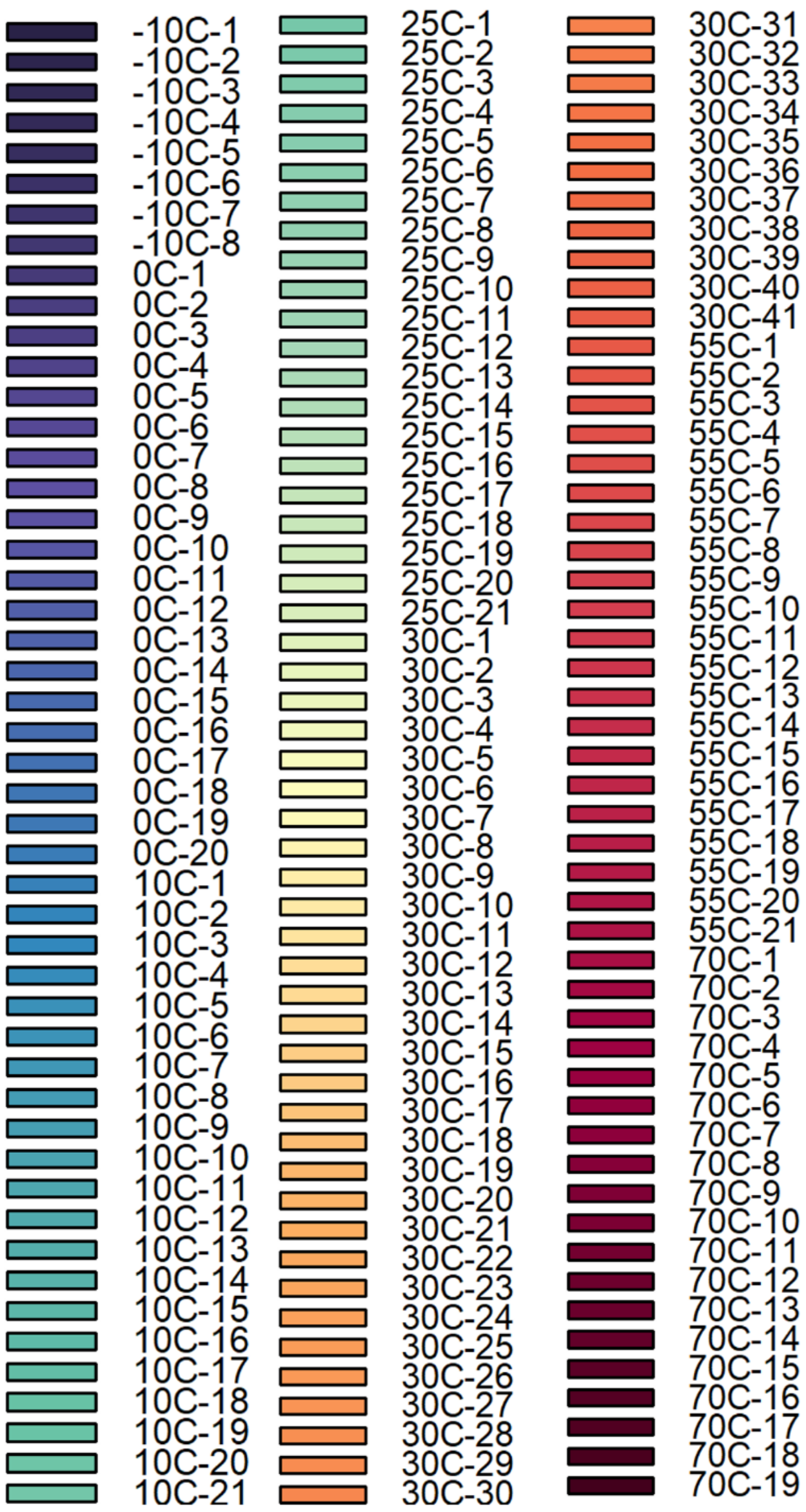

**Supplementary Fig. 17.**

The related battery labels related to the Fig. 3a.

**Supplementary Table 1.**

Model metrics for the battery lifespan prediction results through our method and the other methods[6,7]. (MAE: Mean Absolute Error; RMSE: Root Mean Square Error)

| Name | TRAIN MAPE | TEST MAPE | TRAIN RMSE | TEST RMSE |
|---|---|---|---|---|
| **Training set mean** | 67.54 ± 5 | 75.34 ± 14 | 203.13 ± 12 | 201.07 ± 29 |
| $\mathbf{log_{10}}\left(\boldsymbol{IQR}(\Delta\boldsymbol{Q}_{\mathbf{100-10}}(\boldsymbol{V}))\right)$ | 67.10 ± 5 | 76.12 ± 15 | 201.65 ± 13 | 200.96 ± 29 |
| $\mathbf{log_{10}}\left(\boldsymbol{percentile}(\Delta\boldsymbol{Q}_{\mathbf{100-10}}(\boldsymbol{V}))\right)$ | 66.98 ± 5 | 76.09 ± 15 | 201.45 ± 13 | 200.76 ± 29 |
| $\mathbf{log_{10}}(\Delta\boldsymbol{Q}_{\mathbf{100-10}}(\boldsymbol{V}=\mathbf{4}\boldsymbol{V}))$ | 60.44 ± 4 | 66.42 ± 12 | 198.08 ± 12 | 194.97 ± 30 |
| **Severson et al. variance model** | 66.87 ± 5 | 75.99 ± 15 | 201.37 ± 13 | 200.56 ± 29 |
| **Severson et al. discharge model** | 59.78 ± 4 | 77.01 ± 15 | 194.52 ± 12 | 205.54 ± 28 |
| **Severson et al. full model** | 25.46 ± 2 | 32.43 ± 5 | 133.60 ± 10 | 156.37 ± 24 |
| **Ridge regression on** $\Delta\boldsymbol{Q}_{\mathbf{100-10}}(\boldsymbol{V})$ | 62.92 ± 16 | 80.39 ± 21 | 190.73 ± 45 | 232.73 ± 132 |
| **Elastic net on** $\Delta\boldsymbol{Q}_{\mathbf{100-10}}(\boldsymbol{V})$ | 30.85 ± 4 | 48.82 ± 9 | 129.90 ± 15 | 169.92 ± 26 |
| **PCR on** $\Delta\boldsymbol{Q}_{\mathbf{100-10}}(\boldsymbol{V})$ | 40.22 ± 5 | 48.28 ± 6 | 152.92 ± 15 | 164.13 ± 31 |
| **PLSR on** $\Delta\boldsymbol{Q}_{\mathbf{100-10}}(\boldsymbol{V})$ | 79.90 ± 9 | 87.70 ± 16 | 225.77 ± 19 | 225.13 ± 29 |
| **Random forest on** $\Delta\boldsymbol{Q}_{\mathbf{100-10}}(\boldsymbol{V})$ | 11.94 ± 5 | 40.33 ± 7 | 67.95 ± 22 | 166.82 ± 29 |
| **MLP on** $\Delta\boldsymbol{Q}_{\mathbf{100-10}}(\boldsymbol{V})$ | 16.47 ± 2 | 59.85 ± 12 | 88.39 ± 12 | 225.83 ± 29 |
| **CNN on** $\Delta\boldsymbol{Q}_{(\boldsymbol{V})}$ | 0.42 ± 0 | 29.57 ± 5 | 2.26 ± 1 | 153.16 ± 29 |
| **Random forest on our feature** | 9.28 ± 1 | **26.88 ± 4** | 62.39 ± 7 | **136.96 ± 27** |

**Supplementary Table 2**

Description of the feature operators about automatically polynomial-scale feature space.

| Operator | Description | Type |
|---|---|---|
| **VQ_c** | Interpolate V_c(Voltage) with Q_c(Capacity) | Intra-Cycle Signals |
| **VQ_d** | Interpolate V_d(Voltage) with Q_d(Capacity) | Intra-Cycle Signals |
| **QV_c** | Interpolate Q_c(Capacity) with V_c(Voltage) | Intra-Cycle Signals |
| **QV_d** | Interpolate Q_d(Capacity) with V_d(Voltage) | Intra-Cycle Signals |
| **dVdQ_c** | dV/dQ in charging state(Delta_v means Voltage in odd index minus Voltage in even index of a cycle. Delta_Q is the same, then dV divided by dQ.) | Intra-Cycle Signals |
| **dVdQ_d** | dV/dQ in discharging state | Intra-Cycle Signals |
| **I_c** | Interpolate I_c(Current) with t_c(Time in seconds) | Intra-Cycle Signals |
| **I_d** | Interpolate I_d(Current) with t_d(Time in seconds) | Intra-Cycle Signals |
| **V_c** | Interpolate V_c(Voltage) with t_c(Time in seconds) | Intra-Cycle Signals |
| **V_d** | Interpolate V_d(Voltage) with t_d(Time in seconds) | Intra-Cycle Signals |
| **E_c** | Interpolate E_d(Energy) with t_d(Time in seconds) | Intra-Cycle Signals |
| **E_d** | Interpolate E_d(Energy) with t_d(Time in seconds) | Intra-Cycle Signals |
| **W_c** | Interpolate W_c(Power) with t_d(Time in seconds) | Intra-Cycle Signals |
| **W_d** | Interpolate W_d(Power) with t_d(Time in seconds) | Intra-Cycle Signals |
| **Nan_min** | Minimum value of data, ignoring any NaNs. | Aggregators |
| **Nan_max** | Maximum value of data, ignoring any NaNs. | Aggregators |
| **Nan_mean** | Mean value of data, but it ignores any NaN values in the dataset. | Aggregators |
| **nanskew** | Skewness of data | Aggregators |
| **nankurtosis** | Kurtosis of data | Aggregators |
| **nan_var** | Variance of data | Aggregators |
| **identity** | Raw values, not transformed | Activators |
| **abs** | Absolute value of data | Activators |
| **Signal(a/b)** | Divide all Intra-Cycle Signals like VQ_c in a cycle into groups b and get ath group of them. (b>=a) | Intra-cycle-group operators |
| **Cycle(a/b)** | Divide all training cycles into groups b and get ath group of them. (b>=a) | Intra-cycle-group operators |
| **Cycle(a/b) - Cycle(c/d)** | Subtract value of Cycle(c/d) data from Cycle(a/b) data | Inter -cycle-group operators |

**Supplementary Table 3**

The most salient features for the battery lifespan prediction based on their importance of selection in random forest model.

| Id | Features (Predicting Remaining Useful Life) | Importance score |
|---|---|---|
| **0** | identity(nanmax(Cycle(6/7))[nanvar(VQ_d(1/4))]) | 0.041011 |
| **1** | abs(nanmean(Cycle(3/7) - Cycle(6/7))[nanmean(VQ_d(3/4))]) | 0.027796 |
| **2** | abs(nanmax(Cycle(6/7))[nanvar(VQ_d(1/4))]) | 0.027792 |
| **3** | identity(nanmean(Cycle(4/7) - Cycle(6/7))[nanmean(VQ_d(3/4))]) | 0.02096 |
| **4** | identity(nanmean(Cycle(3/7) - Cycle(6/7))[nanmean(VQ_d(3/4))]) | 0.020626 |
| **5** | identity(nanmax(Cycle(2/7) - Cycle(7/7))[nanmean(VQ_d(1/4))]) | 0.020349 |
| **6** | identity(nanmax(Cycle(5/7))[nanmean(V_d(4/4))]) | 0.015144 |
| **7** | abs(nanmin(Cycle(4/7))[nanmean(V_d(3/4))]) | 0.014298 |
| **8** | abs(nanmin(Cycle(3/7) - Cycle(7/7))[nanmin(QV_c(3/4))]) | 0.014227 |
| **9** | abs(nanmin(Cycle(5/7))[nanmax(V_d(2/4))]) | 0.014209 |
| **10** | identity(nanmax(Cycle(4/7) - Cycle(6/7))[nanmean(VQ_d(3/4))]) | 0.013964 |
| **11** | identity(nanmean(Cycle(3/7) - Cycle(6/7))[nanmax(dVdQ_d(4/4))]) | 0.013888 |
| **12** | abs(nanmax(Cycle(2/7) - Cycle(7/7))[nanmean(VQ_d(2/4))]) | 0.013876 |
| **13** | abs(nanmax(Cycle(6/7))[nanmax(V_d(2/4))]) | 0.013721 |
| **14** | identity(nanmean(Cycle(3/7) - Cycle(6/7))[nanmax(VQ_d(3/4))]) | 0.008297 |
| **15** | identity(nanvar(Cycle(5/7) - Cycle(7/7))[nanmean(VQ_d(3/4))]) | 0.007802 |
| **16** | identity(nanvar(Cycle(5/7) - Cycle(7/7))[nanmax(V_d(1/4))]) | 0.007614 |
| **17** | abs(nanmean(Cycle(4/7))[nanmax(V_d(4/4))]) | 0.0076 |
| **18** | identity(nankurtosis(Cycle(2/7) - Cycle(7/7))[nanmax(dVdQ_d(4/4))]) | 0.007537 |
| **19** | abs(nanmin(Cycle(4/7) - Cycle(7/7))[nanvar(VQ_d(1/4))]) | 0.007458 |

**Supplementary Table 4**

The most salient features for the battery "Knee point" prediction based on their importance of selection in random forest model.

| Id | Features (Predicting "Knee point") | Importance |
|---|---|---|
| **0** | identity(nanvar(Cycle(1/5) - Cycle(3/5))[nanmax(VQ_d(1/4))]) | 0.017690 |
| **1** | abs(nanmean(Cycle(1/5))[nanmean(QV_c(3/4))]) | 0.015220 |
| **2** | identity(nanmean(Cycle(1/5) - Cycle(5/5))[nanvar(VQ_d(4/4))]) | 0.008858 |
| **3** | identity(nanmax(Cycle(1/5))[nanmean(VQ_d(3/4))]) | 0.008854 |
| **4** | identity(nanvar(Cycle(1/5) - Cycle(2/5))[nanmax(VQ_d(1/4))]) | 0.008791 |
| **5** | identity(nanmean(Cycle(1/5))[nankurtosis(E_c(1/4))]) | 0.008336 |
| **6** | abs(nanmax(Cycle(1/5))[nanmax(VQ_c(2/4))]) | 0.008333 |
| **7** | abs(nanmean(Cycle(4/5))[nanvar(VQ_d(1/4))]) | 0.008333 |
| **8** | identity(nanmin(Cycle(2/5))[nanmin(V_d(3/4))]) | 0.008333 |
| **9** | identity(nanmean(Cycle(3/5))[nanskew(VQ_d(1/4))]) | 0.008328 |
| **10** | abs(nanmax(Cycle(4/5) - Cycle(5/5))[nanvar(QV_d(2/4))]) | 0.008319 |
| **11** | identity(nanmean(Cycle(1/5))[nanmean(dVdQ_d(3/4))]) | 0.008293 |
| **12** | identity(nanvar(Cycle(3/5))[nanmin(QV_d(1/4))]) | 0.008269 |
| **13** | identity(nanmax(Cycle(5/5))[nanmin(VQ_c(2/4))]) | 0.008269 |
| **14** | abs(nanmax(Cycle(4/5))[nanmin(VQ_c(2/4))]) | 0.008263 |
| **15** | identity(nanmean(Cycle(2/5) - Cycle(5/5))[nanmin(VQ_c(4/4))]) | 0.008241 |
| **16** | identity(nanmin(Cycle(2/5))[nanmax(V_d(3/4))]) | 0.008241 |
| **17** | identity(nanmax(Cycle(2/5) - Cycle(3/5))[nanmin(VQ_c(2/4))]) | 0.008125 |
| **18** | identity(nanmin(Cycle(2/5))[nanmin(VQ_c(2/4))]) | 0.007843 |
| **19** | identity(nanmean(Cycle(5/5))[nanvar(dVdQ_c(1/4))]) | 0.007843 |

## Supplementary Table 5

Group 0 (HT-LL):
Cluster center:

| Li1s | C1s | O1s | F1s | P2p | Ni3p1 | Co3p1 | Mn3p |
|---|---|---|---|---|---|---|---|
| 21.366923 | 32.98 | 24.451846 | 14.800154 | 4.492154 | 1.037846 | 0.785077 | 0.037538 |

| data_tag | life | temperature | Li1s | C1s | O1s | F1s | P2p | Ni3p1 | Co3p1 | Mn3p | CO | PF | CO/(CO+PF) |
|---|---|---|---|---|---|---|---|---|---|---|---|---|---|
| 25C-19 | 159 | 25 | 20.766 | 33.630 | 27.310 | 12.786 | 3.676 | 1.086 | 0.720 | 0.026 | 60.940 | 16.462 | 0.787074 |
| 25C-20 | 106 | 25 | 18.774 | 33.524 | 25.534 | 15.128 | 4.958 | 1.016 | 1.034 | 0.036 | 59.058 | 20.086 | 0.746375 |
| 70C-3 | 509 | 70 | 20.656 | 36.706 | 21.818 | 13.618 | 5.070 | 0.866 | 0.736 | 0.090 | 58.524 | 18.688 | 0.758241 |
| 70C-7 | 501 | 70 | 22.730 | 30.240 | 21.812 | 17.832 | 5.454 | 0.926 | 0.684 | 0.108 | 52.052 | 23.286 | 0.690711 |
| 70C-9 | 353 | 70 | 21.806 | 32.080 | 20.896 | 17.800 | 4.944 | 1.432 | 1.000 | 0.038 | 52.976 | 22.744 | 0.698596 |
| 70C-12 | 486 | 70 | 23.240 | 31.106 | 23.276 | 15.752 | 5.088 | 0.830 | 0.656 | 0.054 | 54.382 | 20.840 | 0.722738 |
| 30C-30 | 169 | 30 | 21.200 | 33.828 | 25.690 | 13.634 | 3.888 | 0.998 | 0.730 | 0.036 | 59.518 | 17.522 | 0.772685 |
| 30C-31 | 213 | 30 | 19.842 | 33.572 | 24.728 | 15.534 | 4.610 | 1.112 | 0.602 | 0.006 | 58.300 | 20.144 | 0.742968 |
| 30C-32 | 182 | 30 | 18.864 | 35.810 | 24.670 | 13.884 | 4.822 | 1.038 | 0.874 | 0.032 | 60.480 | 18.706 | 0.763888 |
| 30C-41 | 157 | 30 | 20.538 | 37.384 | 27.028 | 9.944 | 2.852 | 1.408 | 0.848 | 0.000 | 64.412 | 12.796 | 0.834954 |
| 55C-1 | 674 | 55 | 23.240 | 29.914 | 24.134 | 16.772 | 4.384 | 0.860 | 0.648 | 0.054 | 54.048 | 21.156 | 0.718212 |
| 55C-11 | 450 | 55 | 22.126 | 31.708 | 25.418 | 14.430 | 4.456 | 1.010 | 0.848 | 0.006 | 57.126 | 18.886 | 0.751465 |
| 55C-15 | 427 | 55 | 23.988 | 29.238 | 25.560 | 15.288 | 4.196 | 0.910 | 0.826 | 0.002 | 54.798 | 19.484 | 0.737058 |

Group 1 (MT-MLL):
Cluster center:

| Li1s | C1s | O1s | F1s | P2p | Ni3p1 | Co3p1 | Mn3p |
|---|---|---|---|---|---|---|---|
| 27.5848 | 22.3544 | 31.771067 | 12.7716 | 3.142067 | 0.935067 | 1.4208 | 0.057067 |

| **data_tag** | **life** | **temperature** | **Li1s** | **C1s** | **O1s** | **F1s** | **P2p** | **Ni3p1** | **Co3p1** | **Mn3p** | **CO** | **PF** | **CO/(CO+PF)** |
|---|---|---|---|---|---|---|---|---|---|---|---|---|---|
| 25C-6 | 54 | 25 | 25.632 | 22.430 | 33.292 | 14.270 | 2.740 | 1.194 | 0.442 | 0.000 | 55.722 | 17.010 | 0.765812 |
| 25C-8 | 71 | 25 | 28.522 | 23.562 | 31.874 | 10.866 | 3.562 | 0.736 | 0.866 | 0.010 | 55.436 | 14.428 | 0.793673 |
| 25C-9 | 76 | 25 | 30.752 | 19.670 | 33.116 | 11.604 | 2.852 | 0.950 | 1.018 | 0.034 | 52.786 | 14.456 | 0.783842 |
| 25C-15 | 89 | 25 | 27.940 | 23.550 | 29.922 | 11.518 | 2.897 | 0.392 | 4.202 | 0.152 | 53.472 | 15.126 | 0.780715 |
| 25C-16 | 118 | 25 | 24.318 | 23.108 | 31.066 | 13.952 | 3.306 | 0.292 | 3.840 | 0.120 | 54.174 | 17.258 | 0.758460 |
| 25C-21 | 85 | 25 | 26.264 | 23.000 | 32.824 | 13.698 | 2.610 | 1.244 | 0.354 | 0.008 | 55.824 | 16.308 | 0.772745 |
| 30C-22 | 191 | 30 | 25.892 | 25.384 | 30.852 | 13.140 | 3.200 | 0.936 | 0.550 | 0.044 | 56.236 | 16.340 | 0.775328 |
| 30C-24 | 248 | 30 | 28.310 | 20.810 | 33.898 | 12.106 | 3.166 | 0.888 | 0.792 | 0.026 | 54.708 | 15.272 | 0.781909 |
| 30C-26 | 143 | 30 | 29.548 | 19.968 | 33.644 | 11.716 | 3.168 | 0.834 | 1.104 | 0.014 | 53.612 | 14.884 | 0.783181 |
| 30C-27 | 163 | 30 | 27.540 | 20.980 | 32.414 | 13.548 | 3.894 | 1.010 | 0.580 | 0.034 | 53.394 | 17.442 | 0.753802 |
| 30C-28 | 172 | 30 | 25.548 | 22.748 | 32.390 | 13.820 | 3.678 | 1.054 | 0.698 | 0.066 | 55.138 | 17.498 | 0.759323 |
| 30C-29 | 305 | 30 | 28.962 | 23.120 | 29.858 | 12.670 | 3.168 | 1.168 | 0.990 | 0.062 | 52.978 | 15.838 | 0.769274 |
| 30C-33 | 854 | 30 | 29.396 | 20.956 | 31.742 | 11.438 | 2.522 | 0.542 | 3.326 | 0.080 | 52.698 | 13.960 | 0.790336 |
| 30C-36 | 233 | 30 | 25.216 | 24.216 | 29.900 | 13.712 | 3.408 | 1.594 | 1.778 | 0.170 | 54.116 | 17.120 | 0.760245 |
| 30C-38 | 439 | 30 | 29.932 | 21.814 | 29.774 | 13.516 | 2.960 | 1.192 | 0.772 | 0.036 | 51.588 | 16.476 | 0.755750 |

Group 2 (LT-SL):
Cluster center:

| **Li1s** | **C1s** | **O1s** | **F1s** | **P2p** | **Ni3p1** | **Co3p1** | **Mn3p** |
|---|---|---|---|---|---|---|---|
| 32.561 | 18.134333 | 25.597333 | 17.761333 | 2.859833 | 0.703 | 2.353667 | 0.042667 |

| **data_tag** | **life** | **temperature** | **Li1s** | **C1s** | **O1s** | **F1s** | **P2p** | **Ni3p1** | **Co3p1** | **Mn3p** | **CO** | **PF** | **CO/(CO+PF)** |
|---|---|---|---|---|---|---|---|---|---|---|---|---|---|
| -10C-1 | 4 | -10 | 34.820 | 13.600 | 21.800 | 24.450 | 2.570 | 0.690 | 2.080 | 0.000 | 35.400 | 27.020 | 0.567126 |
| 25C-1 | 40 | 25 | 32.912 | 18.184 | 23.858 | 19.764 | 3.134 | 0.896 | 1.048 | 0.000 | 42.042 | 22.898 | 0.647777 |

| data_tag | life | temperature | Li1s | C1s | O1s | F1s | P2p | Ni3p1 | Co3p1 | Mn3p | CO | PF | CO/(CO+PF) |
|---|---|---|---|---|---|---|---|---|---|---|---|---|---|
| 25C-3 | 38 | 25 | 32.930 | 19.016 | 25.162 | 17.572 | 2.984 | 1.130 | 1.208 | 0.000 | 44.178 | 20.556 | 0.682161 |
| 25C-4 | 47 | 25 | 32.050 | 19.564 | 28.826 | 13.216 | 2.183 | 0.224 | 4.242 | 0.144 | 48.390 | 15.472 | 0.757593 |
| 25C-5 | 47 | 25 | 32.064 | 19.120 | 27.220 | 14.502 | 2.244 | 0.392 | 4.358 | 0.096 | 46.340 | 16.746 | 0.734476 |
| 30C-25 | 278 | 30 | 30.590 | 19.322 | 26.718 | 17.064 | 4.044 | 0.886 | 1.186 | 0.016 | 46.040 | 21.108 | 0.685623 |

Group 3 (MT-ML):
Cluster center:

| Li1s | C1s | O1s | F1s | P2p | Ni3p1 | Co3p1 | Mn3p |
|---|---|---|---|---|---|---|---|
| 27.033778 | 23.652222 | 25.940889 | 17.484222 | 3.657444 | 0.878444 | 1.561333 | 0.053111 |

| data_tag | life | temperature | Li1s | C1s | O1s | F1s | P2p | Ni3p1 | Co3p1 | Mn3p | CO | PF | CO/(CO+PF) |
|---|---|---|---|---|---|---|---|---|---|---|---|---|---|
| 25C-2 | 42 | 25 | 27.710 | 23.886 | 28.768 | 14.982 | 2.908 | 1.062 | 0.668 | 0.014 | 52.654 | 17.890 | 0.744441 |
| 25C-10 | 57 | 25 | 27.596 | 19.726 | 28.194 | 18.766 | 3.690 | 0.970 | 0.882 | 0.018 | 47.920 | 22.456 | 0.682233 |
| 25C-14 | 98 | 25 | 25.684 | 23.914 | 26.720 | 18.324 | 3.310 | 1.072 | 0.972 | 0.006 | 50.634 | 21.634 | 0.700064 |
| 25C-17 | 121 | 25 | 26.340 | 22.902 | 28.774 | 18.298 | 4.535 | 0.950 | 0.704 | 0.018 | 51.676 | 22.350 | 0.692944 |
| 25C-18 | 125 | 25 | 25.944 | 22.528 | 23.234 | 21.048 | 2.816 | 0.222 | 4.088 | 0.116 | 45.762 | 23.864 | 0.663259 |
| 30C-21 | 251 | 30 | 26.428 | 25.560 | 22.630 | 18.730 | 4.438 | 0.852 | 1.298 | 0.058 | 48.190 | 23.168 | 0.675514 |
| 30C-23 | 303 | 30 | 27.822 | 24.026 | 24.686 | 16.672 | 4.376 | 0.992 | 1.286 | 0.136 | 48.712 | 21.048 | 0.698095 |
| 30C-37 | 164 | 30 | 31.094 | 24.286 | 25.338 | 12.868 | 2.298 | 0.680 | 3.368 | 0.070 | 49.624 | 15.166 | 0.765489 |
| 55C-7 | 683 | 55 | 24.686 | 26.042 | 25.124 | 17.670 | 4.546 | 1.106 | 0.786 | 0.042 | 51.166 | 22.216 | 0.697074 |

Group 4:
Cluster center:

| Li1s | C1s | O1s | F1s | P2p | Ni3p1 | Co3p1 | Mn3p |
|---|---|---|---|---|---|---|---|
| 3.62 | 42.03 | 33.076 | 17.632 | 1.888 | 0.718 | 0.426 | 0.962 |

| data_tag | life | temperature | Li1s | C1s | O1s | F1s | P2p | Ni3p1 | Co3p1 | Mn3p | CO | PF | CO/(CO+PF) |
|---|---|---|---|---|---|---|---|---|---|---|---|---|---|
| 25C-11 | 54 | 25 | 3.62 | 42.03 | 33.076 | 17.632 | 1.888 | 0.718 | 0.426 | 0.962 | 75.106 | 19.866 | 0.789069 |

Group 5:
Cluster center:

| Li1s | C1s | O1s | F1s | P2p | Ni3p1 | Co3p1 | Mn3p |
|---|---|---|---|---|---|---|---|
| 18.0 | 51.434 | 17.158 | 10.156 | 1.656 | 0.636 | 0.862 | 0.098 |

| data_tag | life | temperature | Li1s | C1s | O1s | F1s | P2p | Ni3p1 | Co3p1 | Mn3p | CO | PF | CO/(CO+PF) |
|---|---|---|---|---|---|---|---|---|---|---|---|---|---|
| 70C-14 | 331 | 70 | 18.0 | 51.434 | 17.158 | 10.156 | 1.656 | 0.636 | 0.862 | 0.098 | 68.592 | 11.812 | 0.853065 |

Group 6 (MT-SL):
Cluster center:

| Li1s | C1s | O1s | F1s | P2p | Ni3p1 | Co3p1 | Mn3p |
|---|---|---|---|---|---|---|---|
| 24.794333 | 30.549 | 30.682667 | 9.312333 | 2.505833 | 0.887 | 0.646333 | 0.012 |

| data_tag | life | temperature | Li1s | C1s | O1s | F1s | P2p | Ni3p1 | Co3p1 | Mn3p | CO | PF | CO/(CO+PF) |
|---|---|---|---|---|---|---|---|---|---|---|---|---|---|
| 25C-7 | 47 | 25 | 25.876 | 28.410 | 29.002 | 11.718 | 3.572 | 0.634 | 0.786 | 0.000 | 57.412 | 15.290 | 0.790284 |
| 25C-13 | 60 | 25 | 24.776 | 29.060 | 27.328 | 13.894 | 3.204 | 1.038 | 0.700 | 0.002 | 56.388 | 17.098 | 0.766761 |
| 30C-34 | 203 | 30 | 26.626 | 27.026 | 34.146 | 8.412 | 2.178 | 1.004 | 0.604 | 0.002 | 61.172 | 10.590 | 0.852432 |
| 30C-35 | 201 | 30 | 26.040 | 27.564 | 31.456 | 10.922 | 2.400 | 0.884 | 0.576 | 0.052 | 59.020 | 13.189 | 0.818959 |
| 30C-39 | 238 | 30 | 23.748 | 35.920 | 31.082 | 5.920 | 1.872 | 0.892 | 0.572 | 0.000 | 67.002 | 7.792 | 0.895892 |
| 30C-40 | 246 | 30 | 21.700 | 35.314 | 31.082 | 5.008 | 1.809 | 0.870 | 0.640 | 0.016 | 66.396 | 6.988 | 0.90765 |

Group 7 (HT-LRL):
Cluster center:

| Li1s | C1s | O1s | F1s | P2p | Ni3p1 | Co3p1 | Mn3p |
|---|---|---|---|---|---|---|---|
| 25.456 | 25.8792 | 19.6432 | 22.3996 | 4.3652 | 0.9764 | 1.2448 | 0.0356 |

| data_tag | life | temperature | Li1s | C1s | O1s | F1s | P2p | Ni3p1 | Co3p1 | Mn3p | CO | PF | CO/(CO+PF) |
|---|---|---|---|---|---|---|---|---|---|---|---|---|---|
| 25C-12 | 12 | 25 | 19.912 | 26.866 | 24.132 | 23.768 | 3.454 | 1.142 | 0.676 | 0.048 | 50.998 | 27.222 | 0.651598 |
| 70C-2 | 692 | 70 | 28.216 | 22.586 | 15.548 | 26.150 | 4.564 | 0.634 | 2.224 | 0.086 | 38.134 | 30.714 | 0.553442 |
| 70C-5 | 523 | 70 | 25.504 | 28.520 | 18.758 | 20.440 | 4.874 | 0.770 | 1.116 | 0.016 | 47.278 | 25.314 | 0.649729 |
| 70C-17 | 344 | 70 | 26.114 | 27.994 | 20.122 | 18.994 | 4.694 | 1.028 | 1.044 | 0.008 | 48.116 | 23.688 | 0.670004 |
| 55C-19 | 397 | 55 | 27.534 | 23.430 | 19.656 | 22.646 | 4.240 | 1.308 | 1.164 | 0.020 | 43.086 | 26.886 | 0.613232 |

**Supplementary Table 6**

The most salient features for the battery interfacial chemistry prediction based on their importance of selection in random forest model.

| | Features (Predicting SEI Component Ratios After Etching) | Importance |
|---|---|---|
| 0 | abs(nankurtosis(Cycle(2/5))[nanmax(E_c(2/4))]) | 0.001440 |
| 1 | abs(nanmax(Cycle(1/5) - Cycle(5/5))[nanmax(W_d(4/4))]) | 0.001440 |
| 2 | abs(nanmin(Cycle(1/5))[nanvar(VQ_d(2/4))]) | 0.001432 |
| 3 | abs(nankurtosis(Cycle(2/5))[nanskew(VQ_d(4/4))]) | 0.001402 |
| 4 | identity(nanmean(Cycle(2/5) - Cycle(3/5))[nanskew(W_c(3/4))]) | 0.001357 |
| 5 | identity(nankurtosis(Cycle(1/5) - Cycle(5/5))[nanmin(W_d(1/4))]) | 0.001337 |
| 6 | abs(nanmean(Cycle(4/5))[nanmax(E_d(3/4))]) | 0.001330 |
| 7 | abs(nanmax(Cycle(1/5))[nanmin(E_d(2/4))]) | 0.001322 |
| 8 | identity(nanvar(Cycle(4/5) - Cycle(5/5))[nanskew(W_d(1/4))]) | 0.001315 |
| 9 | identity(nanmin(Cycle(1/5) - Cycle(3/5))[nanmean(QV_c(3/4))]) | 0.001286 |
| 10 | identity(nanskew(Cycle(1/5) - Cycle(2/5))[nanskew(I_d(1/4))]) | 0.001274 |
| 11 | identity(nanmin(Cycle(4/5) - Cycle(5/5))[nanmean(W_c(3/4))]) | 0.001274 |
| 12 | identity(nanvar(Cycle(4/5))[nanskew(I_c(1/4))]) | 0.001266 |
| 13 | abs(nanmin(Cycle(2/5) - Cycle(3/5))[nanskew(W_d(1/4))]) | 0.001236 |
| 14 | identity(nanskew(Cycle(2/5) - Cycle(4/5))[nanvar(I_c(4/4))]) | 0.001235 |
| 15 | identity(nanmax(Cycle(1/5))[nankurtosis(V_d(3/4))]) | 0.001231 |
| 16 | abs(nankurtosis(Cycle(3/5) - Cycle(4/5))[nanmean(E_c(3/4))]) | 0.001228 |
| 17 | identity(nanvar(Cycle(2/5))[nanmin(V_d(1/4))]) | 0.001218 |
| 18 | identity(nankurtosis(Cycle(1/5) - Cycle(4/5))[nanmean(W_d(3/4))]) | 0.001211 |
| 19 | abs(nanvar(Cycle(1/5) - Cycle(5/5))[nanmin(dVdQ_c(3/4))]) | 0.001209 |

**Ref:**

1. Karabasoglu, O. & Michalek, J. Influence of driving patterns on life cycle cost and emissions of hybrid and plug-in electric vehicle powertrains. *Energy Policy* **60**, 445-461 (2013).
2. Yang, X.-G., Liu, T. & Wang, C.-Y. Thermally modulated lithium iron phosphate batteries for mass-market electric vehicles. *Nature Energy* **6**, 176-185 (2021).
3. Fu, R., Wang, H. & Zhao, W. Dynamic driver fatigue detection using hidden Markov model in real driving condition. *Expert Systems with Applications* **63**, 397-411 (2016).
4. Diao, W., Saxena, S., Han, B. & Pecht, M. Algorithm to Determine the Knee Point on Capacity Fade Curves of Lithium-Ion Cells. *Energies* **12**, 2910 (2019).
5. Lloyd, S. Least squares quantization in PCM. *IEEE transactions on information theory* **28**, 129-137 (1982).
6. Attia, P.M., Severson, K.A. & Witmer, J.D. Statistical learning for accurate and interpretable battery lifetime prediction. *Journal of The Electrochemical Society* **168**, 090547 (2021).
7. Severson, K.A., *et al.* Data-driven prediction of battery cycle life before capacity degradation. *Nature Energy* **4**, 383-391 (2019).